\documentclass[11pt]{article}
\usepackage{amssymb,amsmath,epsfig}
\def\one{{{{\rm 1} \kern -.19em {\rm l}}}}
\def\C{{{{\rm {\mbox{\small l}}} \kern -.50em {\rm C}}}}
\def\R{{{{\rm l} \kern -.15em {\rm R}}}}
\def\N{{{{\rm l} \kern -.15em {\rm N}}}}
\def\E{{{{\rm l} \kern -.15em {\rm E}}}}
\def\P{{{{\rm l} \kern -.15em {\rm P}}}}
\def\Z{{{{\rm Z} \kern -.35em {\rm Z}}}}
\def\1{{{{\rm 1} \kern -.35em {\rm 1}}}}

\begin{document}
\begin{sloppypar}
\vspace*{0cm}
\begin{center}
{\setlength{\baselineskip}{1.0cm}{ {\large{\bf
DARBOUX PARTNERS OF HEUN-CLASS POTENTIALS FOR THE 
TWO-DIMENSIONAL MASSLESS DIRAC EQUATION
\\}} }}
\vspace*{1.0cm}
{\large{\sc{Axel Schulze-Halberg$^\ast$}} and {\sc{Artur M. Ishkhanyan$^{\dagger \ddagger}$}}}
\end{center}
\noindent \\
$\ast$ Department of Mathematics and Actuarial Science and Department of Physics, Indiana University Northwest, 3400 Broadway,
Gary IN 46408, USA, E-mail: axgeschu@iun.edu \\ \\
$\dagger$ Russian-Armenian University, Yerevan, 0051 Armenia \\ \\
$\ddagger$ Institute for Physical Research, NAS of Armenia, Ashtarak, 0203 Armenia
\vspace*{.5cm}
\begin{abstract}
\noindent
We apply the Darboux transformation to construct new exactly-solvable cases of the 
two-dimensional massless Dirac equation for potential classes of Lambert-W and inverse exponential type. 
Both of these classes originate from the Heun equation. 
Conditions are devised for transformed potentials to be real-valued, and to be in terms of elementary functions.

\end{abstract}


\noindent \\

\section{Introduction}
The Darboux transformation is a mathematical procedure that provides a mapping between differential 
equations. While the original concept of this transformation \cite{darboux} applied to linear, second-order 
equations only, in the meantime extended versions were constructed that can handle many linear as well 
as nonlinear equations \cite{gu} \cite{matveev}. As such, the primary use of the Darboux transformation 
consists in generating equations, solutions of which can be given in closed form rather than through 
approximative methods. This is particularly important in applications, where differential equations govern the 
behavior of systems that appear in nature: closed-form solutions of such equations can be used in a much more 
versatile way as compared to their approximative counterparts that were obtained through numerical solvers. 
One of the most popular application fields for the Darboux transformation is Quantum Mechanics, in particular, 
quantum systems that are governed by Schr\"odinger or Dirac equations. For example, it was found that the Darboux 
transformation for Schr\"odinger systems is mathematically equivalent to the supersymmetry formalism \cite{bagrov} 
\cite{djsusy}, which is the principal method for the generation of exactly-solvable nonrelativistic quantum systems. 
Due to the vast amount of literature on the topic we refer the reader to the comprehensive review \cite{djtrend} and 
references therein. Besides the nonrelativistic context, a Darboux transformation was also developed for the 
relativistic Dirac equation \cite{nieto} \cite{ekat}. Especially the massless case of the latter equation has been 
subject to intensive research due to its applicability in the area of Dirac materials. This term stands for 
lattice systems featuring quasi-free charge carriers, which at low energy behave like Dirac fermions \cite{cayssol}. 
While graphene is a particularly famous example for a Dirac material \cite{geim} \cite{kats}, other examples are 
d-wave superconductors \cite{bala}, and topological insulators \cite{wehling}, among many others. As indicated above, 
the behavior of low-energy charge carriers in Dirac materials is described accurately by the massless Dirac 
equation. Due to importance of Dirac materials it is an essential task to track down exactly-solvable cases of the 
governing Dirac equation. Even though such cases have been identified and discussed in a variety of contexts, see for example 
\cite{radial} \cite{down0} \cite{hartmann3} and references therein, there are still very few systems known that admit 
closed-form solutions. The purpose of the present work is 
precisely the generation of new exactly-solvable cases to the massless Dirac equation by means of the 
Darboux transformation. To this end, we exploit a link between the latter Dirac equation and a Schr\"odinger-like 
counterpart, thus enabling us to construct an initial Dirac solution from the Schr\"odinger context. Furthermore, 
we focus on two classes of potentials that only recently were found to render the associated 
Schr\"odinger equation exactly-solvable in terms of Heun functions that degenerate to hypergeometric form 
\cite{arturexp}-\cite{artur2}. As a side 
remark let us note that a distinctive feature of these systems consists in hypergeometric 
bound-state Schr\"odinger solutions that do not degenerate to elementary (polynomial) form. Furthermore we point out that the shapes 
of the above potentials are an asymmetric step or a singular well, such that they are 
particularly relevant to the context of Dirac materials. After constructing closed-form hypergeometric solutions of 
our initial Dirac equation, we apply the Darboux transformation in order to generate new exactly-solvable Dirac 
systems. The remainder of this paper is organized as follows: in section 2 we give a short review on decoupling the 
massless Dirac equation in two dimensions, as well as on the Darboux transformation. Section 3 is devoted to 
our first potential class, characterized by the Lambert-W function. In particular, we identify elementary cases of the 
solutions to this potential, discuss properties of the Darboux-transformed potential, and apply our 
Darboux transformation for several parameter settings. In section 4 we follow the same 
procedure for the second potential class that is given in terms of exponential functions.

\section{Preliminaries}
We will now briefly summarize a particular decoupling process for the two-dimensional, stationary Dirac 
equation. Furthermore, a first-order Darboux transformation for the latter Dirac equation is reviewed. To begin with, 
let us introduce the stationary massless Dirac equation in two dimensions that we focus on in this work. We write it 
in the form
\begin{eqnarray}
-i~\sigma_1~\Psi_x(x,y)-i~\sigma_2~\Psi_y(x,y) + \left[U_0(x)-E \right]~\Psi(x,y) &=& 0. \label{dirac}
\end{eqnarray}
Here, indices denote partial differentiation, $\sigma_1$, $\sigma_2$ stand for the Pauli matrices, the potential 
$U_0=(U_0)_{ij}$, $i,j=1,2$, is given by a 2 $\times$ 2 matrix, and the parameter $E$ represents the stationary energy. 

\subsection{Decoupling process}
Since the matrix potential $U_0$ in our Dirac equation (\ref{dirac}) does not depend on the variable $y$, 
we can achieve decoupling by setting
\begin{eqnarray}
\Psi(x,y) &=& \exp\left(i~k_y~y\right) \left[\Psi_a(x), \Psi_b(x) \right]^T, \label{psi} 
\end{eqnarray}
where the wave number $k_y$ describes free motion in the $y$-direction, and the scalar components 
$\Psi_a$, $\Psi_b$ are redefined in terms of functions $\Psi_1$, $\Psi_2$ as follows
\begin{eqnarray}
\Psi_a(x) &=& \Psi_1(x) + \Psi_2(x) \label{psia} \\[1ex]
\Psi_b(x) &=& \Psi_1(x)-\Psi_2(x). \label{psib}
\end{eqnarray}
Assume that the matrix $U_0$ represents a scalar potential, that is, we have $U_0=\mbox{diag}(u_0)$ for a function 
$u_0=u_0(x)$. Substitution of (\ref{psia}), (\ref{psib}) into the Dirac equation (\ref{dirac}) delivers the following constraints
\begin{eqnarray}
\Psi_1''(x) \hspace{-0.1cm} &+& \hspace{-0.1cm} \left\{ \left[u_0(x)-E\right]^2-k_y^2+i~u_0'(x)\right\} \Psi_1(x) ~=~ 0 \label{sse} \\[1ex]
\Psi_2(x) &=& -\frac{i}{k_y} ~\Big\{
i~\Psi_1'(x)-\left[E-u_0(x) \right] \Psi_1(x)\Big\}, \label{psi2}
\end{eqnarray}
where the prime denotes differentiation. Alternatively, the second component (\ref{psi2}) can also be defined as 
a solution of 
\begin{eqnarray}
\Psi_2''(x) +\left\{ \left[u_0(x)-E\right]^2-k_y^2-i~u_0'(x)\right\} \Psi_2(x) ~=~ 0 \label{psi2alt} 
\end{eqnarray}
In summary, once a solution $\Psi_1$ of the second-order equation (\ref{sse}) 
is known, we can generate a corresponding solution to our Dirac equation (\ref{dirac}). More precisely, we 
first calculate $\Psi_2$ from (\ref{psi2}), and afterwards plug both $\Psi_1$ and $\Psi_2$ into (\ref{psia}),(\ref{psib}). This 
determines a two-component solution (\ref{psi}) of the Dirac equation (\ref{dirac}).

\subsection{Darboux transformation for the Dirac equation} 
We will now adapt results from \cite{ekat} to the present Dirac equation (\ref{dirac}) in order to construct a first-order 
Darboux transformation. To this end, let $u$ be an invertible 2 $\times$ 2 matrix given by
\begin{eqnarray}
u(x) &=&  
\left(
\begin{array}{lll}
\Psi_a(x)_{|k_y = \lambda_0} & \Psi_a(x)_{|k_y = \lambda_1} \\[1ex]
\Psi_b(x)_{|k_y = \lambda_0} & \Psi_b(x)_{|k_y = \lambda_1}
\end{array}
\right), \label{u}
\end{eqnarray}
where $\Psi_a$, $\Psi_b$ were defined in (\ref{psi}), and $\lambda_0$, $\lambda_1$ stand for 
arbitrary constants. We will refer to $u$ as transformation matrix and to $\lambda_0$, $\lambda_1$ as 
transformation parameters. The Darboux transformation of the solution 
$\Psi$ to our initial Dirac equation (\ref{dirac}) is then given by
\begin{eqnarray}
\Phi(x,y) &=& u(x) \left[ u(x)^{-1} \Psi(x,y) \right]_x. \label{darboux}
\end{eqnarray}
This two-component function solves the following Dirac equation that is the transformed counterpart of (\ref{dirac})
\begin{eqnarray}
-i~\sigma_1~\Phi_x(x,y)-i~\sigma_2~\Phi_y(x,y) + \left[U_1(x)-E \right]~\Phi(x,y) &=& 0, \label{diract}
\end{eqnarray}
where the transformed Darboux partner potential matrix $U_1$ is given by
\begin{eqnarray}
U_1(x) &=& U_0(x) -\sigma_2 \left[\sigma_3,~u'(x)~u(x)^{-1} \right], \label{v1}
\end{eqnarray}
note that the prime denotes differentiation, and that $[\cdot,\cdot]$ represents the commutator. Furthermore, 
substitution of the Pauli matrices and (\ref{u}) shows that $U_1$ is diagonal if $U_0$ is diagonal. Before we 
conclude this short review, let us add a comment on the preservation of boundary conditions and solution 
asymptotics through (\ref{darboux}). In general, asymptotics of the transformed solution depends strongly on the choice of 
transformation parameters $\lambda_0$ and $\lambda_1$ in the matrix (\ref{u}). However, if these 
parameters are chosen such that all entries in the latter matrix as well as the initial solution components 
satisfy a particular asymptotic condition, then the transformed solution will obey this condition as well. 
We will make use of this property when we generate bound states in our example sections further below.

\subsection{Motivation and applicability in physics}
Apart from the general comments that were made in the introduction regarding applicability of the present work, 
in this paragraph we will provide a more specific motivation for the problem that we are considering, and how it 
relates to existing applications in physics. We are pursuing here two principal objectives.
\begin{itemize}
\item {\bf{Presentation and study of two new exactly-solvable Dirac models:}} We will present and discuss 
properties of two potentials for which the massless, two-dimensional Dirac equation admits closed-form 
solutions. Both of these potentials can take the form of an asymmetric barrier, the shape of which can be 
varied through parameters. As such, these potentials and the closed-form solutions we discuss here 
can be particularly interesting for applications related to electron confinement in Dirac materials. In fact, 
potentials of barrier form and related Dirac systems appear in a wide variety of applications, see e.g. 
\cite{allain} \cite{downingzero} \cite{erementchouk} \cite{jakubsky}, just to mention a few. Having closed-form 
solutions at one's disposal facilitates the study of the underlying systems considerably. Since this is particularly true 
when the aforementioned solutions are elementary (meaning that they do not contain any special functions), 
we will elaborate on conditions, under which the solutions to our two Dirac sytems take elementary form. 
\item {\bf{Generation of further exactly-solvable Dirac models:}} Application of the Darboux transformation to 
the two systems that we will focus on in this work, leads to new Dirac potentials that allow for closed-form solutions of the Dirac equation. 
Being deformed versions of their initial counterparts, our transformed potentials maintain their barrier form while 
exhibiting an additional feature such as a finite-height spike. As such, they offer additional features for e.g. modeling 
electron-confining barriers within a Dirac material. The analysis and application of such systems is facilitated if 
closed-form or even elementary solutions are at hand. 
\end{itemize}

\section{The Lambert-W Dirac system}
In this section we will focus on the Dirac equation (\ref{dirac}) for a specific potential that can be expressed 
by means of Lambert-W functions. This potential can take two qualitatively different forms, both of which 
render our Dirac equation exactly-solvable. In the following we will give a brief discussion of the potential and 
the associated Dirac solution, afterwards identify elementary subcases, and in the final part apply our 
Darboux transformation to generate new exactly-solvable potentials of Lambert-W type for the Dirac equation.

\subsection{Potential and Dirac solution}
\begin{figure}[h]
\begin{center}
\epsfig{file=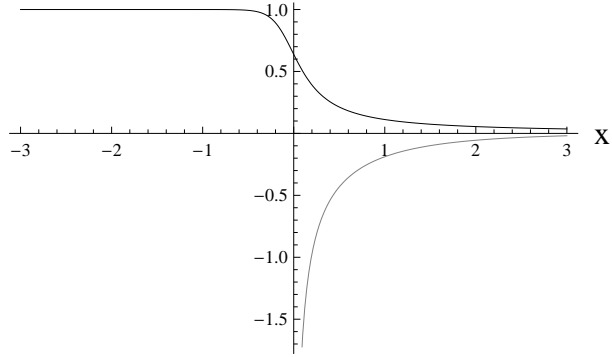,width=8cm}
\caption{Graphs of the initial potential (\ref{pot}) for parameter settings $x_0=V_0=0, ~V_1=1, ~\sigma=1/10$ 
(black curve), and $x_0=-1-i \pi,~ V_0=1=-V_1,~ \sigma=-1$ (gray curve).}
\label{fig0}
\end{center}
\end{figure}

The potential that we focus on in this section reads as follows
\begin{eqnarray}
U_0(x) &=& \left\{V_0+\frac{V_1}{1+W \hspace{-.1cm}\left[\exp\left(
\frac{x-x_0}{\sigma}
\right)
\right]
} \right\} I_2. \label{pot}
\end{eqnarray}
Note that here $I_2$ stands for the 2$\times$2 identity matrix, $W$ denotes the Lambert-W function \cite{abram}, and 
$V_1 \neq 0,~V_0,~\sigma$ are real constants. Furthermore, the parameter $x_0$ either takes a real value or is given by 
$x_0=x_1+i \pi \sigma$ for a real constant $x_1$. In the first case, the diagonal component of (\ref{pot}) represents 
an asymmetric smooth step, the geometry of which is determined by the aforementioned parameters. 
In the second case, the diagonal 
component of (\ref{pot}) is an infinite well with a singularity at $x=x_1-\sigma$, defined on a positive or 
negative half-line. Figure \ref{fig0} visualizes the potential for two different parameter settings.
 \noindent
For further details, the reader is referred to \cite{artur} \cite{artur2}. Observe that 
(\ref{pot}) represents a scalar potential for our Dirac equation because it is diagonal and the 
diagonal entries are equal. As described in section 2.1, we can construct a solution to the Dirac equation 
(\ref{dirac}) with potential (\ref{pot}) by providing a solution of the second-order equation (\ref{sse}). In fact, after 
substituting our potential, the latter equation becomes exactly-solvable in terms of confluent hypergeometric 
functions. A solution can be written in the form
\begin{eqnarray}
\Psi_1(x) &=& z^\frac{\gamma}{2}~\exp\left(\frac{\delta}{2}~z\right)~\frac{d}{dz}
\left\{
\exp\left[-\frac{\delta+s_0}{2}~z \right] {}_1F_1\left(\alpha,\gamma,s_0~z \right)
\right\}_{\Big| z = W[\exp(\frac{x-x_0}{\sigma})]}. \label{psi1lam}
\end{eqnarray}
Note that ${}_1F_1$ represents the confluent hypergeometric function \cite{abram}, and that 
the following abbreviations are in use
\begin{eqnarray}
\begin{array}{llllllllll}
K_0 &=& \sqrt{k_y^2-(E-V_0)^2} & & K_1 &=& \sqrt{k_y^2-(E-V_0-V_1)^2} \\[1ex]
\alpha &=& \frac{\sigma}{2~K_0} \left[(K_0+K_1)^2+V_1^2 \right] &
 & \gamma &=& 2~\sigma~K_1 \\[1ex]
\delta &=& 2~\sigma~(K_1-i~V_1) &  & s_0 &=& 2~\sigma~K_0. 
\end{array} \label{lamset}
\end{eqnarray}
Let us point out that (\ref{psi1lam}), along with (\ref{psi2}) does not represent the general solution of the Dirac 
equation, but only a particular case. This can be seen by taking into account that the function $\Psi_1$ is a solution 
to the second-order equation (\ref{sse}). While the general solution of this equation consists of two linearly independent 
particular solutions, in (\ref{psi1lam}) we present one of these solutions. It is possible to construct the general solution 
through reduction of order. This process, however, involves integrals that we can only solve numerically. For this 
reason, we will restrict ourselves here to the particular solution (\ref{psi1lam}). As a consequence of this restriction, 
we will not be able to fully investigate physical properties of systems that we generate through our Darboux 
transformation. It is straightforward to see that the Dirac solution (\ref{psi}) associated with (\ref{psi1lam}) takes a very long and 
involved form: not only does the calculation of $\Psi_2$ through (\ref{psi2}) involve the derivative of (\ref{psi1lam}), 
but the final components (\ref{psia}) and (\ref{psib}) require a linear combination of two already long expressions. 
For this reason we omit to show the explicit form of the latter solution components.

\subsection{Elementary cases}
\begin{figure}[h]
\begin{center}
\epsfig{file=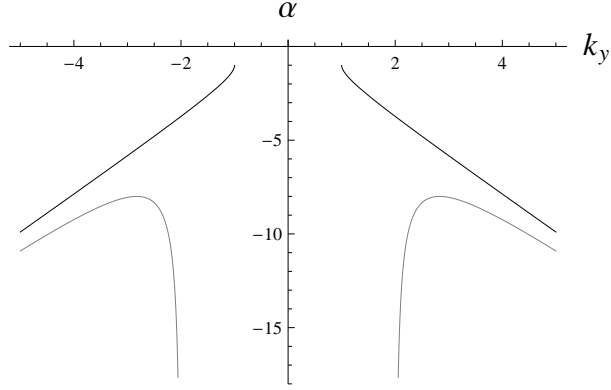,width=8cm}
\caption{Graphs of $\alpha$ dependent on $k_y$, as given in (\ref{alpha}), for the settings 
$\sigma=-1,~V_0=0,~V_1=-1$ (black curve), and $\sigma=-1,~V_0=2,~V_1=-4$ (gray curve).}
\label{fig1}
\end{center}
\end{figure}

The Dirac solution (\ref{psi}) associated with $\Psi_1$ in (\ref{psi1lam}) is a special function, as it contains confluent 
hypergeometric functions as well as the Lambert-W function. It is known that under certain conditions, 
confluent hypergeometric functions become elementary by degenerating to polynomials. In the present section 
we aim to derive such conditions for the confluent hypergeometric functions in (\ref{psi1lam}). The purpose 
of this is twofold. First, we will be able to construct particular Dirac solutions that are elementary, that is, they 
do not contain any confluent hypergeometric or other special functions. Note that the Lambert-W function is 
elementary as the inverse of an elementary function. The 
second purpose of constructing elementary Dirac solutions is related to the Darboux transformation: if we 
use an elementary transformation matrix (\ref{u}), the resulting transformed Dirac potentials will also be 
elementary. Before we start with the construction of conditions for our confluent hypergeometric functions 
in (\ref{psi1lam}) to degenerate, we restrict ourselves for simplicity to the case of zero energy, that is, we 
fix $E=0$. Now, inspection of $\Psi_1$ in (\ref{psi1lam}) reveals that its confluent hypergeometric functions 
degenerate if and only if their first argument equals a nonpositive integer. Let us write this first argument $\alpha$ 
in explicit form by substituting the abbreviations in (\ref{lamset}). We arrive at the representation
\begin{eqnarray}
\alpha &=& \frac{\sigma\left\{V_1^2+\left[\sqrt{k_y^2-V_0^2}+\sqrt{k_y^2-(V_0+V_1)^2}\right]^2\right\}}
{2~\sqrt{k_y^2-V_0^2}}. \label{alpha}
\end{eqnarray}
Following up with our argumentation above, we need this expression to equal a nonpositive integer. 
Consequently, we can write this condition as
\begin{eqnarray}
\frac{\sigma\left\{V_1^2+\left[\sqrt{k_y^2-V_0^2}+\sqrt{k_y^2-(V_0+V_1)^2}\right]^2\right\}}
{2~\sqrt{k_y^2-V_0^2}} &=& -n, ~~~n=0,1,2,...\label{elem}
\end{eqnarray}
Since $V_0,~V_1$, and $\sigma$ are considered fixed parameters of the potential, we can only use $k_y$ 
to satisfy condition (\ref{elem}). In other words, we see $\alpha$ as a function of $k_y$. Furthermore, 
it is important to reiterate that all parameters entering in (\ref{elem}) are real-valued. As this is true particularly for 
$k_y$, we must discard any non-real solution of (\ref{elem}), which can be done by deriving suitable parameter 
restrictions. To start with, we observe that the right side of (\ref{elem}) is negative or zero. Since on the 
left side all terms except for $\sigma$ are defined to be positive or zero, a necessary conditon for having 
real-valued solutions of (\ref{elem}) is 
\begin{eqnarray}
\sigma &<& 0. \label{sigmalam}
\end{eqnarray}
The next parameter constraints we will derive concern the domain and the range of $\alpha$ as a function of $k_y$. 
Reality of all terms on the left side and the observation that $\alpha$ is an even function with respect to $k_y$ 
give the domain as
\begin{eqnarray}
|k_y| ~>~ \max\left\{|V_0|,|V_0+V_1| \right\}. \label{kydomain}
\end{eqnarray}
In order to determine the range of $\alpha$, we can restrict ourselves to positive values $k_y>0$ 
because it is an even function. We first find that $\alpha$ is either a strictly decreasing function or 
it has a single maximum $\alpha_{max}$. These two situations must be handled separately. 
Figure \ref{fig1} shows two graphs of $\alpha$ that we will briefly discuss below for illustration purposes.
 \noindent

\paragraph{Case 1: \boldmath{$\alpha$} is monotonic.} 
In this first case we can find the range to be given by
\begin{eqnarray}
\alpha ~<~ \frac{\sigma~V_1~(V_0+V_1)}{\sqrt{V_1~(2~V_0+V_1)}}, \label{rangenomax}
\end{eqnarray} 
recall that $\sigma$ is negative due to (\ref{sigmalam}). In this case, two solutions are contributed 
for any admissible value of $n$. These solutions are equal in absolute value, but opposite in sign. There is an 
infinite number of admissible values for $n$, since $n$ needs to comply with (\ref{rangenomax}) as
\begin{eqnarray}
-n ~<~ \frac{\sigma~V_1~(V_0+V_1)}{\sqrt{V_1~(2~V_0+V_1)}}. \label{nrange}
\end{eqnarray}
Let us now have a look at figure \ref{fig1}. The black curve in the figure is associated with the parameter values $\sigma=-1,~V_0=0,~V_1=-1$. 
We observe that this curve does not have a maximum. Note that we 
do not take into account the left endpoint of the $k_y$-domain (\ref{kydomain}) because it is an open interval. 
Consequently, the range of 
$\alpha$ is given by (\ref{rangenomax}). Substitution of the present parameter values gives 
$\alpha < -1$, which coincides with the result shown in figure \ref{fig1}. Consequently, if $n$ satisfies this constraint, it can be plugged into condition (\ref{elem}), which in turn delivers 
two solutions for $k_y$. Upon substitution of any of these solutions, (\ref{psi1lam}) becomes an elementary function, 
and so does the resulting two-component solution (\ref{psi}) of the Dirac equation (\ref{dirac}), as 
outlined in 2.1. As an example for such a scenario let us now consider the parameter 
setting corresponding to the black curve in figure \ref{fig1}, that is, we set $\sigma=-1,~V_0=0,~V_1=-1$. In addition, we 
specify the constant $x_0$ that appears in the potential (\ref{pot}) as $x_0=0$. These settings render the latter 
potential as a strictly decreasing step function that tends to zero at negative infinity and to negative one at 
positive infinity. Substitution of our settings into (\ref{nrange}) gives the result $n > 1$, such that the allowed values 
for $n$ are $2,3,4,...$. We can now solve our constraint (\ref{elem}) for the current parameter settings in order to 
obtain values of $k_y$ that render the function (\ref{psi1lam}) as an elementary function. 
Solution of (\ref{elem}) gives the result
\begin{eqnarray}
|k_y| &=& \frac{1}{2} \left( n+\frac{1}{n} \right), ~~~n=2,3,4,... \label{kyx}
\end{eqnarray}
As an example let us substitute $n=2$ into the latter expression, resulting in the positive value $k_y=5/4$. 
We now plug our current parameter settings and $k_y=5/4$ into (\ref{psi1lam}) and (\ref{psi2}). We obtain
\begin{eqnarray}
\Psi_1(x) &=& \left(\frac{50}{3}+i~\frac{25}{3}\right) \exp\left(-\frac{5}{4}~x \right)-
\left(\frac{4}{3}-i\right) \exp\left(-\frac{5}{4}~x \right) \frac{1}{W[\exp(-x)]^2}+ \nonumber \\[1ex]
&+&
\left(10-i ~\frac{10}{3}\right)~\exp\left(-\frac{5}{4}~x \right) \frac{1}{W[\exp(-x)]} \nonumber \\[1ex]
\Psi_2(x) &=& -\frac{5}{3}~\exp\left\{\frac{3}{4}~x+2~W\left[\exp(-x)\right]\right\}
\Bigg\{\hspace{-.2cm}
-i+(2+6~i)~W\left[\exp(-x)\right]+ \nonumber \\[1ex]
&+&(10+5~i)~W\left[\exp(-x)\right]^2\Bigg\}. \nonumber
\end{eqnarray}
These elementary component functions form a solution to the Dirac equation (\ref{dirac}) for the potential (\ref{pot}) 
by means of (\ref{psi})-(\ref{psib}).

\paragraph{Case 2: \boldmath{$\alpha$} has a maximum.} 
In the second case we can find the range by 
determining the maximum's location. Without giving details on this elementary calculation, we state that 
the range is given by $\alpha ~<~ \alpha_{max}$, where we have
\begin{eqnarray}
\alpha_{max} &=& \sigma \left\{2~V_0~[-3~V_0-2~V_1+\sqrt{V_0~(9~V_0+4~V_1)}]\right\}^{-\frac{1}{2}} \times 
\nonumber \\[1ex]
&\times& \Bigg\{
V_1^2+\frac{1}{2}\Bigg[
\sqrt{-3~V_0^2-2~V_0~V_1+V_0~\sqrt{V_0~(9~V_0+4~V_1)}}+ \nonumber \\[1ex]
&+&
\sqrt{-3~V_0^2-6~V_0~V_1-2~V_1^2+\sqrt{V_0^3~(9~V_0+4~V_1)}}
\Bigg]^2
\Bigg\}. \label{alphamax}
\end{eqnarray} 
Therefore, in order to have an admissible value $n$ for our condition (\ref{elem}), this value must be less than 
the maximum of $\alpha$. In other words, $n$ must comply with the constraint
\begin{eqnarray}
-n &<& \alpha_{max}, \label{nalpha}
\end{eqnarray}
where $\alpha_{max}$ is defined in (\ref{alphamax}). Looking at the gray curve in figure \ref{fig1}, we see that it 
has a maximum, thus the range of $\alpha$ is obtained by plugging the parameter values 
into (\ref{alphamax}), leading to $\alpha < -8$. This confirms the graph shown in figure \ref{fig1}. Now, suppose 
that for a particular parameter setting we find from (\ref{nalpha}) an admissible value for $n$ and plug it 
into (\ref{elem}). This condition will then deliver four solutions for $k_y$ that consist of two pairs with opposite signs, but 
equal absolute value. Note that this is in contrast to the previous case of monotonic $\alpha$, where 
(\ref{elem}) delivered only two solutions per fixed value of $n$. Once inserted, each of these solutions for $k_y$ renders 
the function $\Psi_1$ from (\ref{psi1lam}) elementary. As an immediate consequence, the corresponding 
Dirac solution (\ref{psi}) will also consist of elementary functions, as demonstrated in section 2.1 Let us now 
illustrate this process further by providing an example. We choose parameter settings as 
$\sigma=-1/4,~V_0=1/2,~V_1=-1,~x_0=0$. Upon substitution of these settings into (\ref{nalpha}) we obtain 
$n>1/2$, such that the admissible values for $n$ are $n=1,2,3,...$. Now, 
solving equation (\ref{elem}) with respect to $k_y$ yields the four solutions
\begin{eqnarray}
|k_y| &=& \sqrt{2~n^2 \pm n~\sqrt{4~n^2-1}},~~~n=1,2,3,... \label{ky2}
\end{eqnarray}
Here the $\pm$ symbols indicate that for each sign a solution is recovered. 
If we now plug the lowest admissible value $n=1$ into expression (\ref{ky2}), we obtain the positive solutions 
$k_y=\sqrt{2 \pm \sqrt{3}}$. Upon taking $k_y=\sqrt{2-\sqrt{3}}$ and substituting this value into 
(\ref{psi1lam}) and (\ref{psi2}), we obtain the functions $\Psi_1$ and $\Psi_2$ in the form
\begin{eqnarray}
\Psi_1(x) &=&  \frac{1}{4}~\exp\Bigg\{
-\frac{1}{2}~(-2+\sqrt{3})~x-\frac{1}{4}~(-2+\sqrt{3})~W\left[\exp(-4~x)\right]
\Bigg\} ~\times \nonumber \\[1ex]
&\times&
\Bigg\{
-2+i-\sqrt{3}+\Bigg(-2-i+\sqrt{3}\Bigg)~W\left[\exp(-4~x)\right]
\Bigg\} \nonumber \\[1ex]
\Psi_2(x) &=& \Bigg\{
\sqrt{8}+\sqrt{8}~W\left[\exp(-4~x)\right]\Bigg]
\Bigg\}^{-1}~\Bigg\{
(1+\sqrt{3})~\exp\left\{-\frac{1}{4}~(-2+\sqrt{3})~\times \right. \nonumber \\[1ex]
&\times& \left. \Big[2~x+W\left[\exp(-4~x)\right]\Big]\right\}
\Big\{
i-(-2+\sqrt{3})~\exp\left[-8~x-2~W\left[\exp(-4~x)\right] \right]+ \nonumber \\[1ex]
&+& \Big[(2+i)-\sqrt{3}\Big]~W\left[\exp(-4~x)\right]
\Big\}
\Bigg\}. \nonumber
\end{eqnarray}
The next pair of functions $\Psi_1$, $\Psi_2$ is generated by 
plugging $k_y=\sqrt{2+\sqrt{3}}$ into (\ref{psi1lam}) and (\ref{psi2}), giving
\begin{eqnarray} 
\Psi_1(x) &=& \frac{1}{4}~\exp\Bigg[
-\frac{1}{2}~(2+\sqrt{3})~x\Bigg]~W\left[\exp(-4~x)\right]^{-\frac{1}{4} (2+\sqrt{3})}  ~\times \nonumber \\[1ex]
&\times&
\Bigg\{-2+i+\sqrt{3}-(2+i+\sqrt{3})~W\left[\exp(-4~x)\right]
\Bigg\}  \nonumber \\[1ex]
\Psi_2(x) &=& \frac{1}{\sqrt{2}~(5+3~\sqrt{3})}~
\exp\left[-\frac{1}{2}~(2+\sqrt{3})~x \right] W\left[\exp(-4~x)\right]^{-\frac{1}{4} (2+\sqrt{3})}  ~\times \nonumber \\[1ex]
&\times& \Bigg\{
i~(2+\sqrt{3})+(7+4~\sqrt{3})~W\left[\exp(-4~x)\right]
\Bigg\}. \nonumber
\end{eqnarray}
Each pair of these elementary component functions forms a solution of the Dirac equation (\ref{dirac}) for the potential (\ref{pot}) 
through the relations (\ref{psi})-(\ref{psib}). We do not show the corresponding expressions in explicit form 
because of their length.

\subsection{Darboux transformation}
We will now apply the first-order Darboux transformation from section 2.2 to our Dirac equation (\ref{dirac}) for the 
Lambert-W potential (\ref{pot}). To this end, we need to define the functions $\Psi_a$ and $\Psi_b$ in our 
transformation matrix (\ref{u}). These functions are obtained as described in section 2.1: in the first step we 
take $\Psi_1$ from (\ref{psi1lam}), then substitute it into the formulas (\ref{psia}), (\ref{psib}), and (\ref{psi2}). 
Observe that according to section 2.2 the transformed potential (\ref{v1}) is diagonal because its 
initial counterpart (\ref{pot}) is. In the final step of preparing a Darboux transformation we must choose the 
transformation parameters $\lambda_0$ and $\lambda_1$. This choice 
determines if the entries of the transformed Dirac potential matrix (\ref{v1}) are real-valued, and it 
determines if the entries are equal. Since both of these properties are crucial from a physical point of view, we will 
now relate them to our parameters. We have the following statements.
\begin{itemize}
\item Assume that $x_0$ is real-valued. The entries of the transformed potential $U_1$ as given in (\ref{v1}) 
are real-valued if and only if the following conditions are satisfied.
\begin{eqnarray}
\mbox{
Both $\lambda_0$, $\lambda_1$ are real-valued, and $|\lambda_0|>|V_0+V_1|$, $|\lambda_1|>|V_0+V_1|$.}
\label{real1}
\end{eqnarray}
\item Assume that $x_0=x_1+i \sigma \pi$. The entries of the transformed potential $U_1$ as given in (\ref{v1}) 
are real-valued if and only if the following conditions are satisfied.
\begin{eqnarray}
\mbox{
Both $\lambda_0$, $\lambda_1$ are real-valued, and $|\lambda_0|>|V_0+V_1|$, $|\lambda_1|>|V_0+V_1|$, and 
$x_1-\frac{1}{2}<\sigma<0$.} \nonumber \\
\label{real2}
\end{eqnarray}
\item The transformed potential $U_1$ as given in (\ref{v1}) has equal diagonal entries if and only if 
\begin{eqnarray}
\lambda_0 = - \lambda_1. \label{real3}
\end{eqnarray}
\end{itemize}
It is important to point out that the first two of these conditions are not general properties of the 
Darboux transformations, but hold solely for the present system, that is, if the initial potential and the 
transformation matrix are defined as in (\ref{pot}) and (\ref{u}) for (\ref{psi1lam}), respectively. In contrast to this, the 
third condition (\ref{real3}) holds in general, that is, independent of the specific Dirac potential we are considering.
The proof of properties (\ref{real1}) and (\ref{real2}) requires to determine real and imaginary part of the 
transformed potential for all possible parameter settings. This is a very tedious procedure, as the transformed 
potential contains two Wronskians and a regular determinant that do not allow for simplification, see the 
next paragraph for a representation of the transformed potential. For this reason we prefer to omit 
the proof of the aforementioned two properties. We will show now, however, a proof of the third property (\ref{real3}).

\paragraph{Diagonality of the transformed potential matrix.}
In order to analyze the transformed potential 
(\ref{v1}), let us first write the potential difference $U_1-U_0$ in component form. Upon 
introducing transformation matrix components through $u=(u_{ij})$, $i,j=1,2$, evaluation of (\ref{v1}) gives the result
\begin{eqnarray}
U_1(x)-U_0(x) &=& -\sigma_2 \left[\sigma_3,~u'(x)~u(x)^{-1} \right] \nonumber \\[1ex]
&=& -\sigma_2~\sigma_3~u'(x)~u(x)^{-1}-\sigma_2~u'(x)~u(x)^{-1}~\sigma_3 \nonumber \\[1ex]
&=& \frac{2~i}{\det[u(x)]}~
\mbox{diag}\left[W_{u_{21},u_{22}}(x), W_{u_{12},u_{11}}(x)\right], \label{dv}
\end{eqnarray}
where the symbol $W$ denotes the Wronskian of the functions stated in the index. We can say a bit more about 
(\ref{dv}) if we take into account how the components of the transformation 
matrix $u$ relate to solutions of certain equations. Let us first recall from section 2 that the 
pair $(u_{11},u_{21})^T$ and the pair $(u_{12},u_{22})^T$ each form a solution to our Dirac equation (\ref{dirac}). 
Next, we decompose those pairs according to (\ref{psia}), (\ref{psib}), that is, we write
\begin{eqnarray}
u_{11}(x) &=& u_{1,+}(x)+u_{1,-}(x) \qquad \qquad  u_{21}(x) ~=~ u_{1,+}(x)-u_{1,-}(x) \label{upm1} \\[1ex]
u_{12}(x) &=& u_{2,+}(x)+u_{2,-}(x) \qquad \qquad  u_{22}(x) ~=~ u_{2,+}(x)-u_{2,-}(x), \label{upm2}
\end{eqnarray}
introducing functions $u_{1,\pm}$ and $u_{2,\pm}$. Before we continue, let us note that the definitions 
(\ref{upm1}) and (\ref{upm2}) imply
\begin{eqnarray}
\det[u(x)] &=& 2~u_{2,+}(x)~u_{1,-}(x)-2~u_{1,+}(x)~u_{2,-}(x). \label{det}
\end{eqnarray}
Now, following the argumentation from section 2, these 
functions are solutions of the equations
\begin{eqnarray}
u_{1,+}''(x) + \left[ u_0(x)^2-\lambda_0^2+i~u_0'(x)\right] u_{1,+}(x) &=& 0 \label{eqw1} \\[1ex]
u_{1,-}''(x) + \left[ u_0(x)^2-\lambda_0^2-i~u_0'(x)\right] u_{1,-}(x) &=& 0 \label{eqw2} \\[1ex]
u_{2,+}''(x) + \left[ u_0(x)^2-\lambda_1^2+i~u_0'(x)\right] u_{2,+}(x) &=& 0 \label{eqw3} \\[1ex]
u_{2,-}''(x) + \left[ u_0(x)^2-\lambda_1^2-i~u_0'(x)\right] u_{2,-}(x) &=& 0, \label{eqw4}
\end{eqnarray}
recall that in the present case we have set $E=0$. We need these four equations now in order to evaluate the 
Wronskians in (\ref{dv}). More precisely, the derivatives of these Wronskians can be evaluated by 
substituting the latter equations. We find
\begin{eqnarray}
W_{u_{21},u_{22}}'(x) &=& u_{21}''(x)~u_{22}(x)-u_{22}''(x)~u_{21}(x) \nonumber \\[1ex]
&=& (\lambda_1^2-\lambda_0^2)~[u_{1,+}(x)-u_{1,-}(x)]~[u_{2,+}(x)-u_{2,-}(x)]~+~i~u_0'(x)~\det[u(x)] \label{wd1} \\[1ex]
W_{u_{12},u_{11}}'(x) &=& u_{12}''(x)~u_{11}(x)-u_{11}''(x)~u_{12}(x) \nonumber \\[1ex]
&=& (\lambda_0^2-\lambda_1^2)~[u_{1,+}(x)+u_{1,-}(x)]~[u_{2,+}(x)+u_{2,-}(x)]~+~i~u_0'(x)~\det[u(x)], \label{wd2}
\end{eqnarray}
observe that we used our identity (\ref{det}). Inspection of (\ref{wd1}) and (\ref{wd2}) shows that both can only be 
equal if $\lambda_0$ and $\lambda_1$ are equal up to a sign. Since they cannot be the same, as this would render the 
transformation matrix $u$ singular, the only option is $\lambda_0=-\lambda_1$. Upon substituting this into the 
derivatives (\ref{wd1}) and (\ref{wd2}), we obtain
\begin{eqnarray}
W_{u_{21},u_{22}}'(x)_{\mid \lambda_0=-\lambda_1} &=& W_{u_{12},u_{11}}'(x)_{\mid \lambda_0=-\lambda_1} 
~=~i~u_0'(x)~\det[u(x)]. \nonumber
\end{eqnarray}
We conclude that the Wronskians must be the same up to a constant. This constant is zero, as can be seen by 
substituting $\lambda_0=-\lambda_1$ into our equations (\ref{eqw1})-(\ref{eqw4}): since in this case the pairs 
(\ref{eqw1}), (\ref{eqw3}) and (\ref{eqw2}), (\ref{eqw4}) become equal, our definitions (\ref{upm1}) and (\ref{upm2}) imply 
that the Wronskians $W_{u_{21},u_{22}}$ and $W_{u_{12},u_{11}}$ are the same. Consequently, the diagonal 
components of the potential difference in (\ref{dv}) are equal and we have the representation
\begin{eqnarray}
\left[U_1(x)-U_0(x)\right]_{\mid \lambda_0=-\lambda_1} 
&=& -\frac{2}{\mbox{det}[u(x)]} \left\{\int\limits^x u_0'(t)~\mbox{det}[u(t)]~dt \right\} I_2, \label{potrep}
\end{eqnarray}
where $I_2$ stands for the $2 \times 2$ identity matrix. This representation implies reality of the potential 
difference, as long as the determinant of the transformation matrix and the initial potential are real-valued. However, 
the representation refers solely to the particular case $\lambda_0=-\lambda_1$. In general, proving the 
model-specific reality conditions (\ref{real1}) and (\ref{real2}) requires to identify parameter-dependent 
real and imaginary part of the functions contained in (\ref{dv}). As mentioned above, for this reason we omit to 
show the associated proof here, but proceed with examples.

\begin{figure}[h]
\begin{center}
\epsfig{file=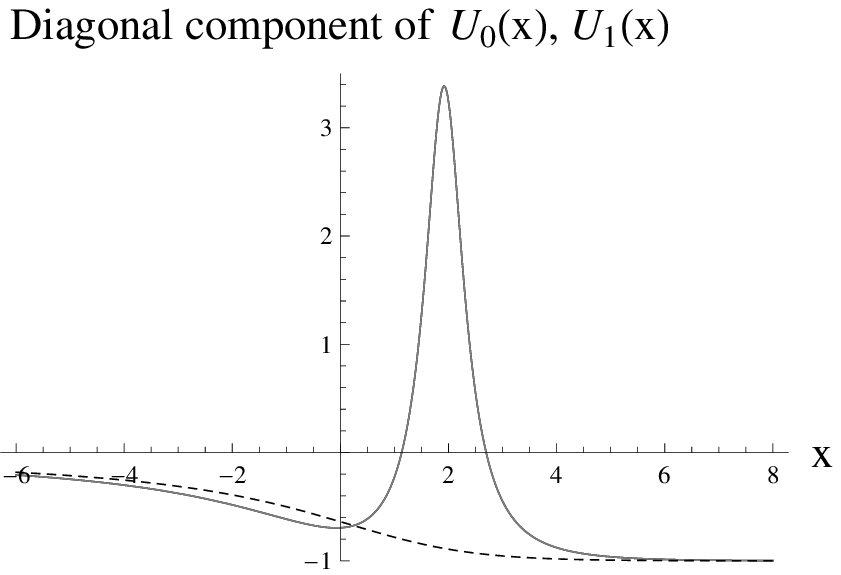,width=8cm}
\caption{Graphs of the transformed potential (\ref{u1x}) (solid curve) and its initial counterpart (\ref{pot}) 
(dashed curve) for the parameter settings (\ref{setx1}), $\lambda_0=-5/4,~\lambda_1=5/4$.}
\label{fig1_1}
\end{center}
\end{figure}

\paragraph{Example: potential step with elementary partner.} In our first application of the Darboux transformation 
we substitute the following parameter settings into our initial potential (\ref{pot}). 
\begin{eqnarray}
V_0~=~0 \qquad \qquad V_1=1\qquad \qquad \sigma~=~-1 \qquad \qquad x_0 ~=~ 0. 
\label{setx1}
\end{eqnarray}
These settings render the diagonal component of the potential matrix as an asymmetric, monotically decreasing 
step function, see the dashed curve in figure \ref{fig1_1}. In order to perform our Darboux transformation, it 
remains to define the transformation parameters $\lambda_0$, $\lambda_1$ that enter in the matrix (\ref{u}). 
In the present example we will choose these parameters such that the Darboux-transformed potential becomes 
real-valued, has equal diagonal elements, and consists of elementary functions only. The first two of these 
properties are governed by conditions (\ref{real1}) and (\ref{real3}), respectively. 
Substitution of our settings (\ref{setx1}) results in the conditions
\begin{eqnarray}
|\lambda_j| ~>~ 1, ~~~j=0,1
 \qquad \mbox{and} \qquad 
\lambda_0~=~ -\lambda_1. \label{realx}
\end{eqnarray}
The last property is attained if our transformation parameters $\lambda_0$, $\lambda_1$ are solutions of 
equation (\ref{elem}) with respect to $k_y$. Following the procedure described in section 3.2, we first 
observe from our settings (\ref{setx1}) that condition (\ref{sigmalam}) is satisfied. Next, we plug these 
settings into (\ref{elem}), which gives
\begin{eqnarray}
-k_y-\sqrt{k_y^2-1} &=& -n. \label{realx1}
\end{eqnarray}
Since the left side is a monotonically decreasing function with respect to $k_y$, we must now find the admissible values 
for $n$ from (\ref{nrange}). Evaluation of the latter condition reveals $n>1$. Upon inserting the lowest 
admissible value $n=2$ into (\ref{realx1}), we obtain $k_y=\pm 5/4$. Hence, the choice 
$\lambda_0=-5/4$ and $\lambda_1=5/4$ complies with both conditions (\ref{realx}) and (\ref{realx1}), such that 
the Darboux transformation will generate a potential (\ref{v1}) that has real-valued and equal entries, given 
in terms of elementary functions. We verify this by plugging (\ref{setx1}) and 
$\lambda_0=-5/4=-\lambda_1$ into the transformation matrix (\ref{u}). Subsequent evaluation of the 
transformed potential (\ref{v1}) gives
\begin{eqnarray}
U_1(x) \hspace{-.2cm} &=& \hspace{-.2cm} \Bigg\{ \hspace{-.1cm} -1+16W\left[\exp(-x)\right]+10W\left[\exp(-x)\right]^2-200W\left[\exp(-x)\right]^3
-125W\left[\exp(-x)\right]^4\Bigg\} \times \nonumber \\[1ex]
&\times& \hspace{-.2cm} 
\Bigg\{1+W\left[\exp(-x)\right]\Bigg\}^{-1}
\Bigg\{
1-12W\left[\exp(-x)\right]+30W\left[\exp(-x)\right]^2+100W\left[\exp(-x)\right]^3+ \nonumber \\[1ex]
&+& \hspace{-.2cm} 125W\left[\exp(-x)\right]^4
\Bigg\}^{-1}I_2. \label{u1x}
\end{eqnarray}
We see that this potential generated by the Darboux transformation satisfies all desired properties, see figure 
\ref{fig1_1} for a visualization.
 \noindent

\begin{figure}[h]
\begin{center}
\epsfig{file=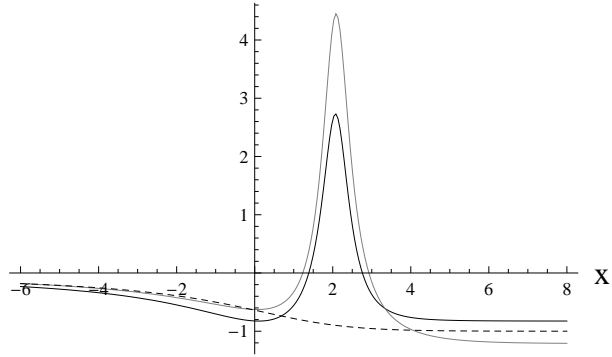,width=8cm}
\caption{The graphs show the diagonal components $(U_1)_{11}$ (black solid curve), 
$(U_1)_{22}$ (gray curve) of the transformed potential (\ref{v1}), and its initial counterpart (\ref{pot}) 
(dashed curve) for the parameter settings (\ref{setx1}), $\lambda_0=-1.2,~\lambda_1=1.35$.}
\label{fig1_2}
\end{center}
\end{figure}

\paragraph{Example: potential step with non-elementary partner.} The second application of our 
Darboux transformation 
is concerned with a slight modification of the previous case. In particular, we maintain the settings (\ref{setx1}), 
which preserves our reality and diagonality conditions (\ref{realx}) and (\ref{realx1}). In contrast to the previous 
example, this time we fix the transformation parameters as $\lambda_0=-1.2$ and $\lambda_1=1.35$. 
We observe that these parameter values comply with the left condition in 
(\ref{realx}), such that the transformed potential (\ref{v1}) will be real-valued. However, we also observe that 
the present choice of parameter values does neither satisfy the right condition in (\ref{realx}) nor condition (\ref{realx1}), 
no matter which admissible value of $n$ we use. As a consequence, the transformed potential matrix (\ref{v1}) 
will have different diagonal entries that are not elementary functions. The explicit 
form of the latter functions can be obtained by substituting (\ref{setx1}) and our transformation 
parameter values into (\ref{v1}). Since this form consists of excessively large expressions in terms of 
hypergeometric functions, we do not display them here. Instead, we show graphs of those functions in 
figure \ref{fig1_2}. Inspection of the figure shows that - as expected - the diagonal entries of the transformed 
potential are real-valued and different.
 \noindent

\begin{figure}[h]
\begin{center}
\epsfig{file=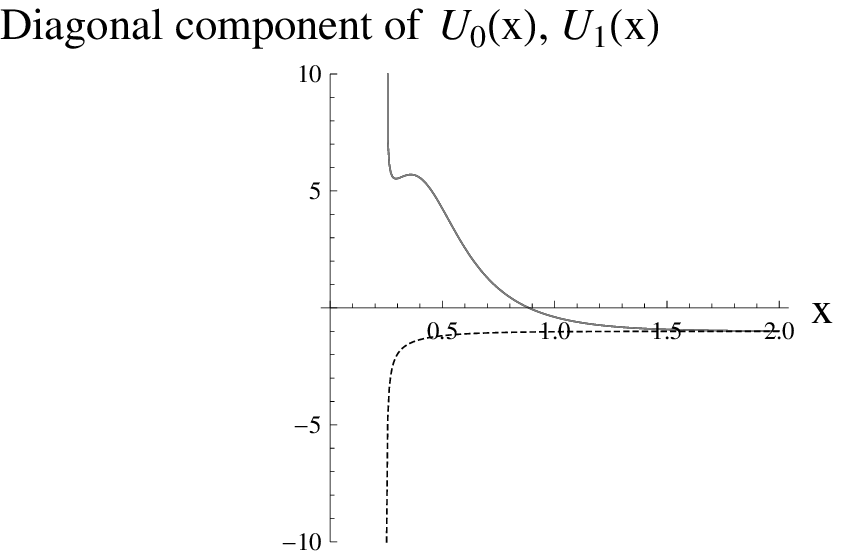,width=8cm}
\caption{Graphs of the transformed potential (\ref{u1xx}) (solid curve) and its initial counterpart (\ref{pot}) 
(dashed curve) for the parameter settings (\ref{setx2}), $\lambda_0=-17/8,~\lambda_1=17/8$.}
\label{fig1_3}
\end{center}
\end{figure}

\paragraph{Example: singular potential with elementary partner.} While the last two applications featured a 
step function in the initial potential, this time we choose a singular version of (\ref{pot}) by means of the 
parameter setting
\begin{eqnarray}
V_0~=~0 \qquad \qquad V_1=-1\qquad \qquad \sigma~=~-\frac{1}{4} \qquad \qquad
x_0 ~=~ -\frac{i~\pi}{4}. \label{setx2}
\end{eqnarray}
According to the short review of our initial potential in section 3.1, the settings (\ref{setx2}) yield a diagonal 
component of (\ref{pot}) that has a singularity at $x=1/4$ and is defined on the half-line $x>1/4$. The dashed curve in 
figure \ref{fig1_3} shows a graph of this function. Before we can perform our Darboux transformation, we must 
now define the transformation parameters $\lambda_0$ and $\lambda_1$ according to the desired properties of 
the transformed potential. Let us require that the latter potential has real-valued, equal diagonal entries that 
are elementary functions. The first two of these properties are handled by means of conditions (\ref{real2}) and 
(\ref{real3}). Upon substitution of our settings (\ref{setx2}) we obtain 
\begin{eqnarray}
|\lambda_j|~>~1,~~~j=0,1 \qquad \mbox{and} \qquad \lambda_0~=~-\lambda_1. \label{realx2}
\end{eqnarray}
In addition to these constraints we must choose our transformation parameters such that the transformed 
potential consists of elementary functions. To this end, we first take into account that the necessary 
condition (\ref{sigmalam}) is met. We can now plug our settings (\ref{setx2}) into (\ref{elem}) that gives 
\begin{eqnarray}
-\frac{k_y}{4}-\frac{1}{4}~\sqrt{k_y^2-1} &=&  -n. \label{realx3}
\end{eqnarray}
The left side of this equation is monotonically decreasing in $k_y$, such that we can find 
admissible values for the integer $n$ from (\ref{nrange}) as $n>1/4$. Upon substituting the lowest admissible 
value $n=1$ into (\ref{realx3}), we obtain the solutions $k_y=\pm17/8$. Consequently, the choice 
$\lambda_0=-17/8=-\lambda_1$ satisfies both conditions (\ref{realx2}) and (\ref{realx3}), such that the 
transformed potential (\ref{v1}) has all the desired properties. In fact, insertion of the above values for our 
transformation parameters and the parameter settings (\ref{setx2}) into our transformed potential (\ref{v1}), 
we obtain
\begin{eqnarray}
U_1(x) &=& \frac{1+34 W\left[-\exp(-4x)\right]+17 W\left[-\exp(-4x)\right]^2}
{1+3 W\left[-\exp(-4x)\right]+19 W\left[-\exp(-4x)\right]^2+17 W\left[-\exp(-4x)\right]^3}~I_2. \label{u1xx}
\end{eqnarray}
We observe that this matrix has equal, real-valued entries that consist of elementary functions. The graph of the 
diagonal component can be found in figure \ref{fig1_3}.
 \noindent

\begin{figure}[h]
\begin{center}
\epsfig{file=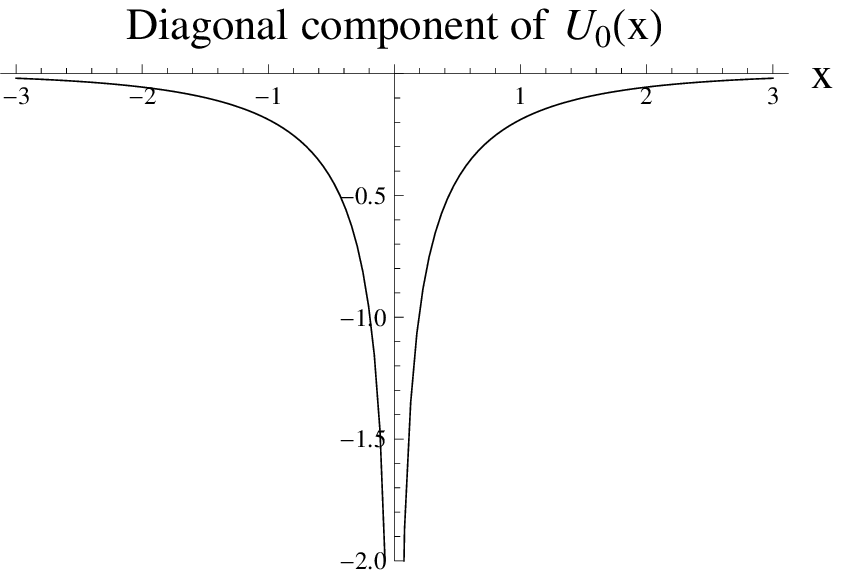,width=7.8cm}
\epsfig{file=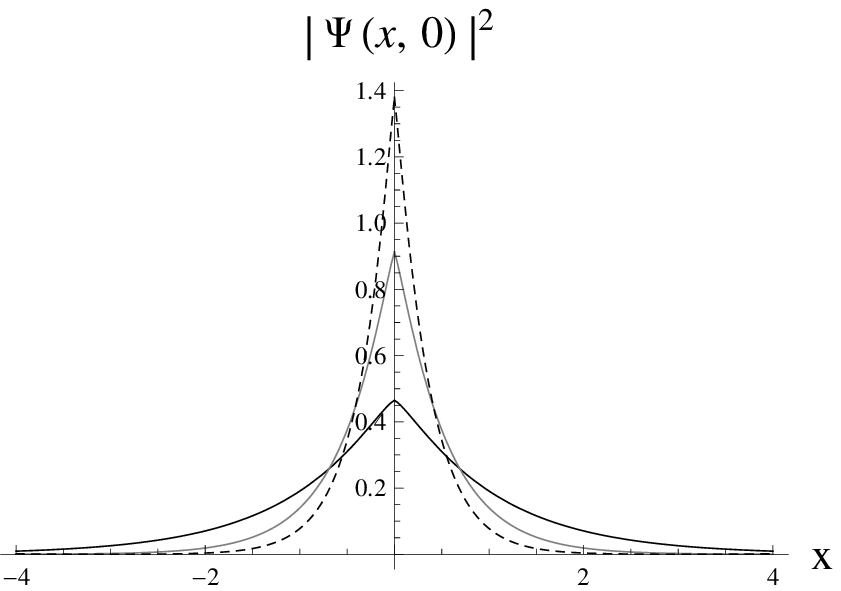,width=7.8cm}
\caption{Left plot: graph of the initial potential (\ref{u0bound}). Right plot: graphs of normalized 
probability densities associated with (\ref{psi1lam}), (\ref{psi2}) for the parameter settings (\ref{setx0}). We have 
$k_y=1/2$ (black solid curve), $k_y=1$ (gray curve), $k_y=3/2$ (dashed curve).}
\label{boundini}
\end{center}
\end{figure}

\begin{figure}[h]
\begin{center}
\epsfig{file=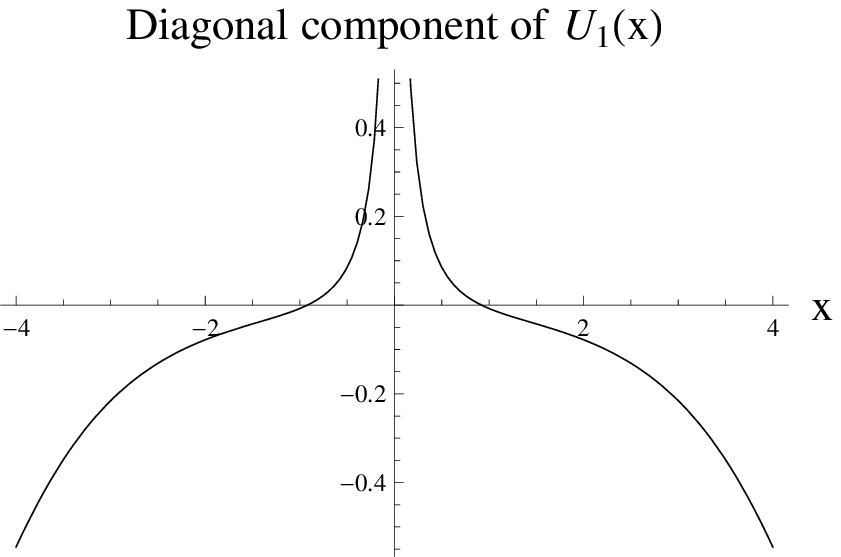,width=7.8cm}
\epsfig{file=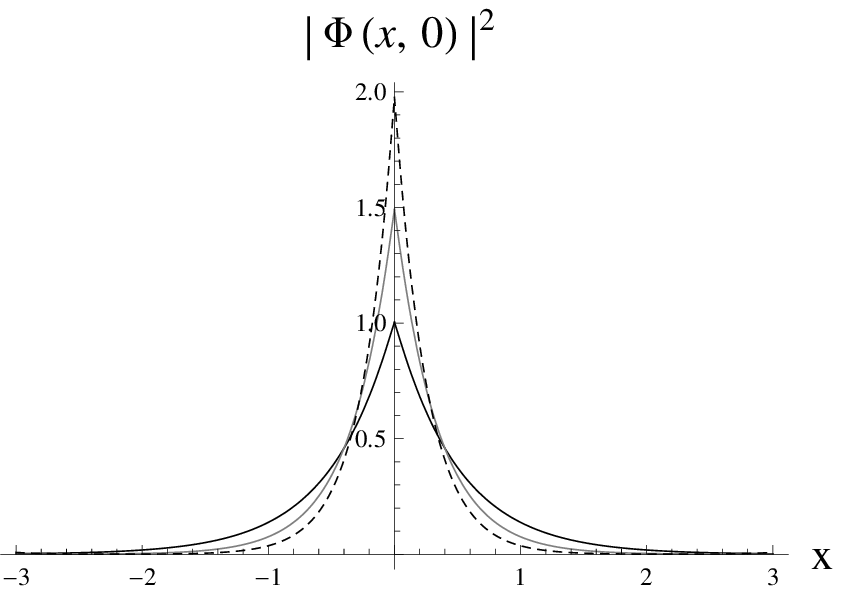,width=7.8cm}
\caption{Left plot: graphs of the transformed potential (\ref{v1}) for parameter settings (\ref{setx0}) and 
$\lambda_0=-\lambda_1=1/2$. Right plot: graphs of normalized 
probability densities associated with (\ref{psi1lam}), (\ref{psi2}) for the parameter settings (\ref{setx0}) 
and $\lambda_0=-\lambda_1=1/2$. We have 
$k_y=1$ (black solid curve), $k_y=3/2$ (gray curve), $k_y=2$ (dashed curve).}
\label{boundfin}
\end{center}
\end{figure}

\paragraph{Example: singular potential admitting bound states.} We will now demonstrate that the singular form of 
our potential (\ref{pot}) can support bound states. To this end, let us first introduce the parameter setting that will 
be used in this example. We define
\begin{eqnarray}
V_0~=~1 \qquad \qquad V_1=-1\qquad \qquad \sigma~=~-1 \qquad \qquad x_0 ~=~ -1-i~\pi. 
\label{setx0}
\end{eqnarray}
While the associated potential (\ref{pot}) is shown in figure \ref{fig0}, in the present case we extend the problem's 
domain to the whole real axis by replacing $x \rightarrow |x|$ in the potential. We will comment on this 
in detail further below. Let us now incorporate the settings 
(\ref{setx0}) and the aforementioned replacement into the Dirac potential (\ref{pot}). We obtain
\begin{eqnarray}
U_0(x) &=& \left\{1-\frac{1}{1+W\left[-\exp\left(-|x|-1 \right)\right]} \right\} I_2. \label{u0bound}
\end{eqnarray} 
A graph of the potential is shown in the left plot of figure \ref{boundini}.
 \noindent
The associated solutions (\ref{psi1lam}) and (\ref{psi2}) of our Dirac equation (\ref{dirac}) are constructed by insertion 
of (\ref{setx0}). The probability density $|\Psi|^2=|\Psi_1|^2+|\Psi_2|^2$ formed by the solutions is found to be 
finite at zero and integrable at the infinities, provided $k_y$ is a multiple of $1/2$. Hence, for these values of $k_y$, 
our system admits bound states, as is illustrated in figure \ref{boundini}. Note that these bound state solutions are 
continuous at the origin. We can now apply our Darboux transformation, where the first task is to choose 
a transformation matrix (\ref{u}) by specifying transformation parameters $\lambda_0$ and $\lambda_1$. 
The key observation for choosing these parameters is that the Darboux transformation (\ref{darboux}) preserves 
the desired behavior of the transformed solutions at the infinities if the entries of the transformation matrix (\ref{u}) 
satisfy them. More precisely, if the entries of the latter matrix vanish at the infinities along with the initial solution 
components and their derivatives, so will the entries of the inverse matrix and its derivative. We make use of this 
observation by taking the transformation parameters $\lambda_0=1/2=-\lambda_1$. 
The latter settings for the transformation parameters as well as (\ref{setx0}) do not yield 
solutions of (\ref{elem}), such that the transformed solutions will not be elementary functions. 
Furthermore, we expect to obtain a real-valued and diagonal transformed potential matrix, since the present 
parameter settings satisfy conditions (\ref{real2}) and (\ref{real3}). Substitution of these settings into (\ref{v1}) and 
into (\ref{darboux}) yield transformed Dirac solutions and an associated matrix potential, the explicit form 
of which we do not show here due to their length. Instead, we refer the reader to figure \ref{boundfin}, where 
both potential and solutions in form of probability densities are displayed.
 \noindent

\section{The exponential Dirac system}
The next Dirac system that we will study here features a potential governed by the inverse square root of an exponential 
expression. Similar to the Lambert-W potential considered in section 3, the present potential can take different 
qualitative forms  \cite{arturepl}, for both of which the Dirac equation admits exact solutions. We will follow the same route as in 
the previous section: after introducing our system, we work out conditions for elementary subcases, and 
afterwards apply the Darboux transformation, thus constructing new exactly-solvable Dirac systems.

\subsection{Potential and Dirac solution} 
\begin{figure}[h]
\begin{center}
\epsfig{file=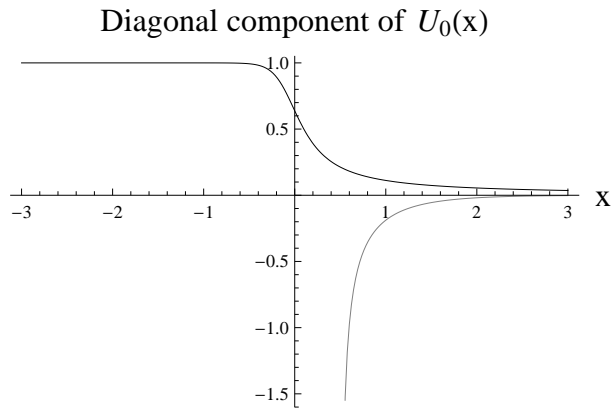,width=8cm}
\caption{Graphs of the initial potential (\ref{potexp}) for parameter settings $x_0=V_0=0, ~V_1=1, ~\sigma=1/10$ 
(black curve), and $x_0=-i \pi/2,~ V_0=1=-V_1,~ \sigma=-1/2$ (gray curve).}
\label{fig2_0}
\end{center}
\end{figure}

Our starting point is the initial potential for our Dirac equation, given by
\begin{eqnarray}
U_0(x) &=& \left\{V_0+\frac{V_1}{\sqrt{1+\exp\left(\frac{x-x_0}{\sigma}\right)}} \right\} I_2. \label{potexp}
\end{eqnarray}
Note that $I_2$ represents the 2$\times$2 identity matrix, $V_1 \neq 0$, $V_0$, $\sigma$ are real-valued 
constants, and $x_0$ either takes a real value or it is given by $x_0=x_1+i \sigma \pi$ for a real $x_1$. 
We observe that the parameter settings are very similar to the Lambert-W potential (\ref{pot}). This similarity 
also holds for the graph of the diagonal matrix entry. More precisely, a real-valued $x_0$ renders the 
graph of the latter 
entry as an asymmetric step function, whereas $x_0=x_1+i \sigma \pi$ results in an infinite well with 
singularity at $x=x_1$. The diagonal component of (\ref{potexp}) is shown in figure \ref{fig2_0} for 
two different parameter settings. 
 \noindent
A more detailed discussion of this potential can be found in \cite{arturexp} \cite{arturepl}. We remark that (\ref{potexp}) 
represents a scalar potential because it is diagonal with equal diagonal entries. Now, recall that if we can 
provide a solution to the Schr\"odinger-type equation (\ref{sse}), then the procedure described in section 2.1 
allows for the construction of an associated solution to the Dirac equation (\ref{dirac}). In fact, a solution 
to (\ref{sse}) can be given in terms of hypergeometric functions. We have
\begin{eqnarray}
\Psi_1(x) &=& (z+1)^{\alpha_1}~(z-1)^{\alpha_2}~\Bigg\{
\frac{q+\beta~z}{2~\gamma-2}~{}_2F_1\left[\alpha,\beta+1,\gamma,\frac{z+1}{2}\right]+ \nonumber \\[1ex]
&+&
{}_2F_1\left[\alpha-1,\beta+1,\gamma-1,\frac{z+1}{2}\right]
\Bigg\}_{\Big| z = \sqrt{1+\exp\left(\frac{x-x_0}{\sigma}\right)}}, \label{psi1exp}
\end{eqnarray}
where ${}_2F_1$ stands for the hypergeometric function \cite{abram}, and we are using the following abbreviations
\begin{eqnarray}
\begin{array}{llllllll}
\alpha_1 &=& \sigma~\sqrt{k_y^2-(E-V_0+V_1)^2} & & \alpha_2 &=& \sigma~\sqrt{k_y^2-(E-V_0-V_1)^2} \\[1ex]
q &=& 2~i~\sigma~V_1-\alpha_1+\alpha_2 & & \gamma &=& 2~\alpha_1+1 \\[1ex]
\beta &=& 2~\sigma~\sqrt{k_y^2-(E-V_0)^2}+\alpha_1+\alpha_2 & & 
\alpha &=& -2~\sigma~\sqrt{k_y^2-(E-V_0)^2}+\alpha_1+\alpha_2.
\end{array} \label{setexp}
\end{eqnarray}
Recall that the corresponding Dirac solution (\ref{psi}) is now constructed by substituting (\ref{psi1exp}) into 
(\ref{psia}), (\ref{psib}), and (\ref{psi2}). Since the resulting expressions are excessively long, we do not 
show them in explicit form.

\subsection{Elementary cases}
The purpose of this section is to track down conditions, under which the solution of our Dirac equation (\ref{dirac}) for the 
potential (\ref{potexp}) consists of elementary functions only. As outlined in section 3.2, elementary cases of 
the present Dirac system are not only interesting in themselves, but also important in regards of generating 
further elementary cases through our Darboux transformation. Let us recall that a sufficient condition for such a 
case to be established is the function $\Psi_1$ in (\ref{psi1exp}) taking elementary form. This occurs if 
either the first or the second argument of the hypergeometric functions in (\ref{psi1exp}) attains a 
negative value or zero. Since these two arguments are governed by the parameters $\alpha$ and $\beta$, 
respectively, we need to consider each of these parameters separately. In contrast to the Lambert-W 
potential discussed in the previous section, there is an additional condition to be considered here.

\begin{figure}[h]
\begin{center}
\epsfig{file=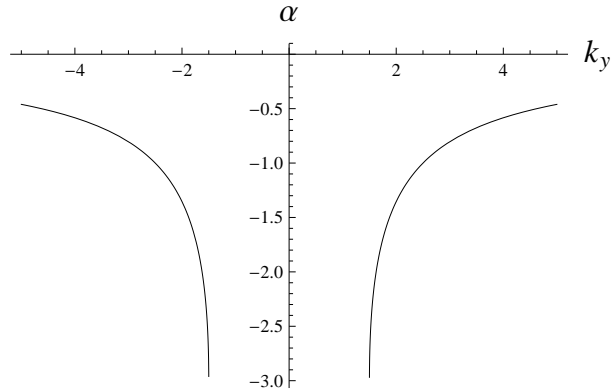,width=8cm}
\caption{Graph of $\alpha$ as a function of $k_y$, as given in (\ref{alphaexp}). The parameter settings are 
$\sigma=1,~V_0=0,~V_1=-3/2$.}
\label{fig2}
\end{center}
\end{figure}

\paragraph{The parameter \boldmath{$\alpha$}.} Let us first combine the settings (\ref{setexp}) 
in order to write our parameter $\alpha$ in explicit form. We find
\begin{eqnarray}
\alpha &=& \sigma~\Bigg[ \hspace{-.1cm} -2~\sqrt{k_y^2-V_0^2}+\sqrt{k_y^2-(V_0+V_1)^2}+\sqrt{k_y^2-(V_0-V_1)^2}
\Bigg]. \label{alphaexp}
\end{eqnarray}
The function (\ref{psi1exp}) becomes elementary if this parameter takes a nonpositive integer value. 
More precisely, we have
\begin{eqnarray}
\sigma~\Bigg[ \hspace{-.1cm} -2~\sqrt{k_y^2-V_0^2}+\sqrt{k_y^2-(V_0+V_1)^2}+\sqrt{k_y^2-(V_0-V_1)^2}
\Bigg] &=& -n,~~~n=0,1,2,... \nonumber \\  \label{alphaexpcon}
\end{eqnarray}
Since $\sigma$, $V_0$ and $V_1$ are fixed parameters asociated with the potential, we must satisfy 
(\ref{alphaexpcon}) by finding appropriate values for $k_y$. In addition, we require that all quantities and terms 
in equation (\ref{alphaexpcon}) are real-valued. Since particularly the square root terms are required to be real-valued, 
we obtain certain restrictions on the parameters that they contain. Before we evaluate these restrictions, let us observe that 
the term in square brackets is always negative or zero, provided we take all square roots to yield nonnegative values. 
Since the left side of (\ref{alphaexpcon}) must be negative or zero, this dictates the constraint
\begin{eqnarray}
\sigma &>& 0. \label{sigmaexp}
\end{eqnarray}
Reality also yields that the left side of (\ref{alphaexpcon}) is subject to the restriction
\begin{eqnarray}
|k_y| &>& \max \left\{|V_0+V_1|,|V_0-V_1| \right\}, \label{kyexp}
\end{eqnarray}
which defines the domain of $\alpha$ with respect to $k_y$. As far as the range of $\alpha$ is concerned, 
a straightforward analysis of (\ref{alphaexp}) reveals that $\alpha$ is an even function that for $k_y>0$ is 
strictly increasing. Its range is given by
\begin{eqnarray}
2~\sigma \left(\sqrt{|V_0~V_1|}-\sqrt{2~|V_0~V_1|+V_1^2}\right) ~<~ \alpha ~<~ 0. \label{alpharange}
\end{eqnarray}
Figure \ref{fig2} shows a graph of $\alpha$ for a particular parameter setting. 
 \noindent
According to (\ref{alpharange}), the parameter $\alpha$ is bounded from below and above. Consequently, 
the number of solutions to our condition (\ref{alphaexpcon}) must be finite. Each solution exists as a pair 
of numbers that have the same absolute value, but opposite sign. Upon substituting any of these solutions for 
$k_y$, the function (\ref{psi1exp}) becomes elementary. As a consequence, it defines an elementary solution 
of the Dirac equation (\ref{dirac}) by means of the relationships (\ref{psi})-(\ref{psi2}). In order to illustrate these 
considerations, let us now look at a particular example, given by the parameter settings 
$\sigma=1,~V_0=0,~V_1=-3/2,~x_0=0$. The task is now to find solutions of (\ref{alphaexpcon}) with respect to 
$k_y$. To this end, we first note that the necessary condition (\ref{sigmaexp}) is satisfied. Next, 
we must first determine admissible values for $n$ from (\ref{alpharange}). Substitution of 
our parameter values gives $-3 < \alpha  <  0$, such that the admissible values 
for $n$ are $n=1$ and $n=2$. Upon inserting the lower value $n=1$, our condition (\ref{alphaexpcon}) 
yields the solutions $k_y = \pm 5/2$. Plugging this value along with our parameter settings into 
(\ref{psi1exp}), and into (\ref{psi})-(\ref{psi2}), returns a solution of our Dirac equation (\ref{dirac}) for the 
potential (\ref{potexp}). This solution reads
\begin{eqnarray}
\Psi(x,y) &=& 
\left(
\begin{array}{cc}
{\displaystyle{\frac{3}{8}~i~\exp\left(2~x+\frac{5}{2}~i~y\right) \sqrt{1+\exp(x)}}} \\[3ex]
{\displaystyle{-\frac{1}{8}~\exp\left(2~x+\frac{5}{2}~i~y\right)}}
\end{array}
\right).
 \nonumber
\end{eqnarray}
As desired, the components of this Dirac solution are given by elementary functions.

\begin{figure}[h]
\begin{center}
\epsfig{file=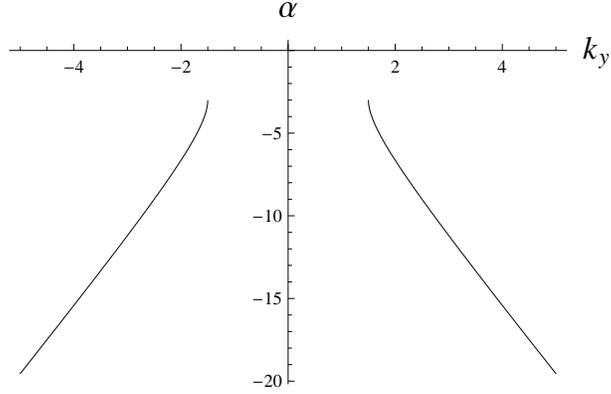,width=8cm}
\caption{Graph of $\beta$ dependent on $k_y$, as given in (\ref{betaexp}), for the settings 
$\sigma=-1,~V_0=0,~V_1=-3/2$.}
\label{fig3}
\end{center}
\end{figure}

\paragraph{The parameter \boldmath{$\beta$}.} In the first step we write the latter parameter in explicit form 
by taking into account (\ref{setexp}). This yields
\begin{eqnarray}
\beta &=& \sigma~\Bigg[ \hspace{-.0cm} 2~\sqrt{k_y^2-V_0^2}+\sqrt{k_y^2-(V_0+V_1)^2}+\sqrt{k_y^2-(V_0-V_1)^2}
\Bigg]. \label{betaexp}
\end{eqnarray}
We remark that this expression differs from (\ref{alphaexp}) by a sign only. Now, the hypergeometric functions 
in (\ref{psi1exp}) degenerate to polynomials if $\beta+1$ equals a nonpositive integer $n$, that is, we have
\begin{eqnarray}
\sigma~\Bigg[ \hspace{-.0cm} 2~\sqrt{k_y^2-V_0^2}+\sqrt{k_y^2-(V_0+V_1)^2}+\sqrt{k_y^2-(V_0-V_1)^2}
\Bigg] &=& -n,~~~n=1,2,3,... \nonumber \\  \label{betaexpcon}
\end{eqnarray}
Taking into account that $n=0$ is not an admissible value here, we must now find real-valued solutions 
of this equation with respect to $k_y$. Let us first observe that the left side of (\ref{betaexpcon}) must be 
negative because the right side is, and because we require all parameters and terms to be real-valued. 
As a consequence, we obtain the constraint
\begin{eqnarray}
\sigma &<& 0. \label{sigmaexp1}
\end{eqnarray}
As (\ref{betaexp}) differs from (\ref{alphaexp}) by a sign only, the left side of (\ref{betaexpcon}) has the 
$k_y$-domain (\ref{kyexp}). Furthermore, it is straightforward to show that the left side of 
(\ref{betaexpcon}) is a strictly decreasing function 
with respect to $k_y$. This observation, together with our reality requirement gives the following range
\begin{eqnarray}
\beta &<& 2~\sigma \left(\sqrt{|V_0~V_1|}+\sqrt{2~|V_0~V_1|+V_1^2}\right). \label{betarange}
\end{eqnarray}
A graph of the function $\beta$ for a particular parameter setting is displayed in figure \ref{fig3}.
 \noindent
As can be seen from (\ref{betarange}), the parameter $\beta$ is unbounded from below. Hence, 
our condition (\ref{betaexpcon}) has an infinite number of solutions, each of which generates a 
corresponding elementary solution of our Dirac equation (\ref{dirac}) through (\ref{psi})-(\ref{psi2}), and 
(\ref{psi1exp}). Each solution occurs as a pair of values that have the same absolute value, but opposite 
signs. We will now look at a brief example for the parameter settings $\sigma=-1,~V_0=0,~V_1=-3/2$, note 
that the necessary condition (\ref{sigmaexp1}) for the existence of solutions to (\ref{betaexpcon}) is 
fulfilled. Now, upon plugging our parameter values into (\ref{betarange}) in order to determine the admissible values for 
$n$, we find $\beta<-3$. Consequently, the allowed values for $n$ are $n=4,5,6,...$. If we insert the lowest value 
$n=4$ into (\ref{betaexpcon}) and solve, we obtain the solutions $k_y = \pm 25/16$. Substituting any of these values 
into (\ref{psi1exp}) and (\ref{psi})-(\ref{psi2}) leads to the Dirac solution
\begin{eqnarray}
\Psi(x,y) &=& 
\left(
\begin{array}{cc}
{\displaystyle{\frac{12}{7}~i~\exp\left(-\frac{9}{16}~x+\frac{25}{16}~i~y \right)~\sqrt{1+\exp(x)}~[13+\exp(x)]}} \\[3ex]
{\displaystyle{
-\frac{3}{7}~\exp\left(-\frac{25}{16}~x+\frac{25}{16}~i~y \right) \bigg\{91+3~\exp(x)~\left[26+\exp(x) \right]
\bigg\}
}}
\end{array}
\right). \nonumber
\end{eqnarray}
The components of this solution are clearly elementary functions as desired.

\paragraph{The parameter \boldmath{$\gamma$}.} The third argument of the hypergeometric function in the solution
(\ref{psi1exp}) can also contribute elementary cases of the latter solution. This is the case if one of the conditions 
$\gamma - \alpha + n = 0$ and $\gamma - \beta - 1 + n = 0$ for a nonnegative integer $n$ is satisfied. 
Upon substituting our settings (\ref{setexp}) into the condition, these two conditions take the form
\begin{eqnarray}
\sigma~\Bigg[ 2~\sqrt{k_y^2-V_0^2}-\sqrt{k_y^2-(V_0+V_1)^2}+\sqrt{k_y^2-(V_0-V_1)^2}\Bigg]+n+1 &=& 0 
\label{gammacon1} \\[1ex]
\sigma~\Bigg[ \hspace{-.1cm}- 2~\sqrt{k_y^2-V_0^2}-\sqrt{k_y^2-(V_0+V_1)^2}+\sqrt{k_y^2-(V_0-V_1)^2}\Bigg]+n+1 &=& 0. 
\label{gammacon2}
\end{eqnarray}
We can analyze these conditions using the results for $\alpha$ and $\beta$ that we obtained in the previous 
paragraphs. In particular, the first condition can yield real solutions only for $\sigma<0$, while the second 
condition can produce real solutions only for $\sigma<0$. In order to avoid repetitive calculations, we omit to state further details here.

\subsection{Darboux transformation}
In this section we will generate new exactly-solvable Dirac systems by applying our Darboux transformation 
to equation (\ref{dirac}) with potential (\ref{potexp}). Since the general procedure is the same as described in 
section 3.3, we will here omit introductory details. Before we perform the Darboux transformation, let us first 
gather some general information. 
\begin{itemize}
\item Assume that $x_0$ is real-valued. The entries of the transformed potential $U_1$ as given in (\ref{v1}) 
are real-valued if and only if the following conditions are satisfied.
\begin{eqnarray}
\mbox{
Both $\lambda_0$, $\lambda_1$ are real-valued solutions of either (\ref{alphaexp}) or (\ref{betaexp}) with 
respect to}~ k_y. \label{real4} 
\end{eqnarray}
In other words, $U_1$ is real-valued if and only if its entries are elementary functions. 
\item Assume that $x_0=x_1+i \sigma \pi$. The entries of the transformed potential $U_1$ as given in (\ref{v1}) 
are real-valued if and only if the following conditions are satisfied.
\begin{eqnarray}
\mbox{
Both $\lambda_0$, $\lambda_1$ are real-valued, and $|\lambda_0|>|V_0|+|V_1|$, $|\lambda_1|>|V_0|+|V_1|$, and 
$\sigma<0$.} \nonumber \\
\label{real5}
\end{eqnarray}
\item The transformed potential $U_1$ as given in (\ref{v1}) has equal diagonal entries if and only if 
\begin{eqnarray}
\lambda_0 = - \lambda_1. \label{real6}
\end{eqnarray}
\end{itemize}
Note that the conditions (\ref{real4}) and (\ref{real5}) are valid only for Darboux transformations performed to the 
Dirac system based on the potential (\ref{potexp}) and the transformation matrix (\ref{u}). The last condition 
(\ref{real6}) holds in a more general context and can be analyzed in the same way as it was done for the 
previous potential (\ref{pot}). In particular, the potential representation (\ref{potrep}) holds in the present case too. 
As in section 3.3, we omit to show the rather 
tedious proof of the above properties (\ref{real4}) and (\ref{real5}), but present examples instead.

\begin{figure}[h]
\begin{center}
\epsfig{file=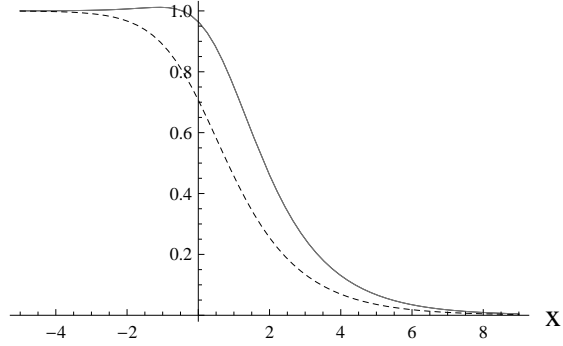,width=8cm}
\caption{Graphs of the transformed potential (\ref{u1x1}) (solid curve) and its initial counterpart (\ref{potexp}) 
(dashed curve) for the parameter settings (\ref{setx3}), $\lambda_0=-13/12,~\lambda_1=13/12$.}
\label{fig2_1}
\end{center}
\end{figure}

\paragraph{Example: step function with elementary partner.} We start out with a potential that has the shape of 
an asymmetric step function. Our parameter settings are as follows
\begin{eqnarray}
V_0~=~0 \qquad \qquad V_1~=~1\qquad \qquad \sigma~=~\frac{3}{4} \qquad \qquad x_0 ~=~ 0. 
\label{setx3}
\end{eqnarray}
The diagonal component of the potential with the latter settings is visualized as the dashed curve in figure \ref{fig2_1}. 
We observe that $x_0$ is a real number, so according to (\ref{real4}) we can generate real-valued potentials only 
if their entries are elementary functions. Since $\sigma$ is positive, we must solve the constraint 
(\ref{alphaexpcon}) in order to obtain suitable values for $k_y$. 
Upon substituting the present settings (\ref{setx3}), 
the latter constraint takes the form
\begin{eqnarray}
-\frac{3}{2}~k_y+\frac{3}{2}~\sqrt{k_y^2-1} &=& -n,~n=0,1,2,... \label{cond1}
\end{eqnarray}
In the next step we need to determine the admissible values for $n$. To this end, we substitute our 
parameter settings (\ref{setx3}) into (\ref{alpharange}), which gives 
\begin{eqnarray}
0 ~<~ n ~<~ \frac{3}{2}-\frac{3~\sqrt{3}}{2} ~\approx~ 1.09808. \nonumber
\end{eqnarray}
Consequently, the only admissible value for $n$ in (\ref{cond1}) is $n=1$. Upon inserting this, we obtain the 
solutions $k_y = \pm 13/12$. We point out that these are the only two values for our settings (\ref{setx3}) that 
lead to elementary transformed Dirac potentials. Hence, let us choose the transformation parameters of our 
Darboux transformation as $\lambda_0=-13/12=-\lambda_1$. Next, we plug these values and the settings 
(\ref{setx3}) into the transformed potential (\ref{v1}). After simplification we find
\begin{eqnarray}
U_1(x) &=& \frac{13+17~\exp\left(\frac{4}{3}~x \right)}
{\left[13+9~\exp\left(\frac{4}{3}~x \right)\right] \sqrt{1+\exp\left(\frac{4}{3}~x \right)}}~I_2. \label{u1x1}
\end{eqnarray}
It is immediate to see that the entries of this potential matrix are elementary functions. Figure \ref{fig2_1} shows 
a graph of the diagonal component, along with the initial partner. 
 \noindent

\begin{figure}[h]
\begin{center}
\epsfig{file=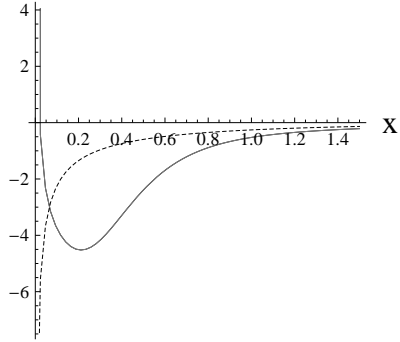,width=8cm}
\caption{Graphs of the transformed potential (\ref{v1}) (solid curve) and its initial counterpart (\ref{potexp}) 
(dashed curve) for the parameter settings (\ref{setx4}), $\lambda_0=-3,~\lambda_1=3$.}
\label{fig2_2}
\end{center}
\end{figure}

\paragraph{Example: singular potential with non-elementary partner.} Let us proceed with our next example that 
is characterized by the parameter settings
\begin{eqnarray}
V_0~=~1 \qquad \qquad V_1=-1\qquad \qquad \sigma~=~-1  \qquad \qquad x_0 ~=~ -i~\pi. 
\label{setx4}
\end{eqnarray}
We observe that $x_0$ is of the form $x_0=x_1+i \pi \sigma$, such that the diagonal component of the 
resulting potential (\ref{potexp}) is defined on the positive half-line and singular at the origin, see the 
dashed curve in figure \ref{fig2_2} for a graph. In addition to the parameter setup (\ref{setx4}), we choose the 
transformation parameters as $\lambda_0=-3=-\lambda_1$. Next, we note that this choice satisfies conditions 
(\ref{real5}) and (\ref{real6}), so we can conclude that the transformed potential (\ref{v1}) has real-valued and 
equal diagonal entries. Furthermore, it is straightforward to verify that condition (\ref{betaexpcon}) is not 
satisfied upon substitution of the above transformation parameters along with the settings (\ref{setx4}). 
This means that the entries of our transformed potential (\ref{v1}) are not elementary functions. 
As a consequence, the explicit form of the latter potential's entries is rather large, such that we omit to state it here. 
Instead, we show a graph in figure \ref{fig2_2}.
 \noindent

\begin{figure}[h]
\begin{center}
\epsfig{file=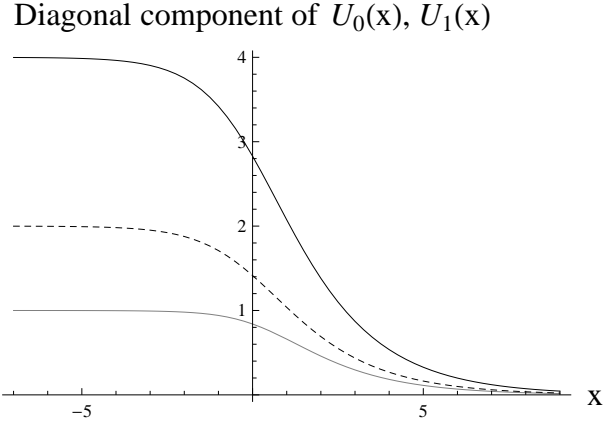,width=8cm}
\caption{The graphs show the diagonal components $(U_1)_{11}$ (black solid curve), 
$(U_1)_{22}$ (gray curve) of the transformed potential (\ref{u1x2}), and its initial counterpart (\ref{potexp}) 
(dashed curve) for the parameter settings (\ref{setx5}), $\lambda_0=-17/4,~\lambda_1=5/2$.}
\label{fig2_3}
\end{center}
\end{figure}

\paragraph{Example: potential step with elementary partner.} In this example we want to use our 
Darboux transformation to generate a potential matrix that consists of real-valued elementary functions, but has 
nonequal diagonal entries. To this end, we employ the parameter settings
\begin{eqnarray}
V_0~=~0 \qquad \qquad V_1=2\qquad \qquad \sigma~=~1 \qquad \qquad x_0 ~=~ 0. 
\label{setx5}
\end{eqnarray}
The desired properties of our transformed potential can all be realized by choosing the transformation 
parameters $\lambda_0$, $\lambda_1$ such that they are solutions of condition (\ref{alphaexpcon}) with 
respect to $k_y$. Note that the necessary condition (\ref{sigmaexp}) for elementary solutions is fulfilled. 
Substitution of our settings (\ref{setx5}) renders the latter equation as
\begin{eqnarray}
-2~k_y+2~\sqrt{k_y^2-4} &=&  -n,~~~n=0,1,2,... \nonumber
\end{eqnarray}
Before we can solve this equation for $k_y$, we must find the admissible values of $n$ by means of 
the range condition (\ref{alpharange}). Taking into account the present parameter values, the latter 
range condition gives $0<n<4$, such that the admissible values for $n$ are $n=1,2,3$. Upon substituting $n=1$ and 
$n=2$ into (\ref{alphaexpcon}), we obtain the solutions $k_y=\pm 17/4$ and $k_y= \pm 5/2$,
respectively. Let us set $\lambda_0=-17/4$ and $\lambda_1=5/2$. Upon inserting these values and (\ref{setx5}) 
into the transformed potential (\ref{v1}), we obtain
\begin{eqnarray}
U_1(x) &=& \left(
\begin{array}{cc}
{\displaystyle{\frac{4}{\sqrt{1+\exp(x)}}}} & 0 \\[1ex]
0 & {\displaystyle{\frac{8+11~\exp(x)}{8 \left[1+\exp(x)
\right]^\frac{3}{2}}}}
\end{array}
\right). \label{u1x2}
\end{eqnarray}
We observe that the diagonal entries of this potential are real-valued and elementary, but not equal. They 
are visualized in figure \ref{fig2_3}. 
 \noindent

\begin{figure}[h]
\begin{center}
\epsfig{file=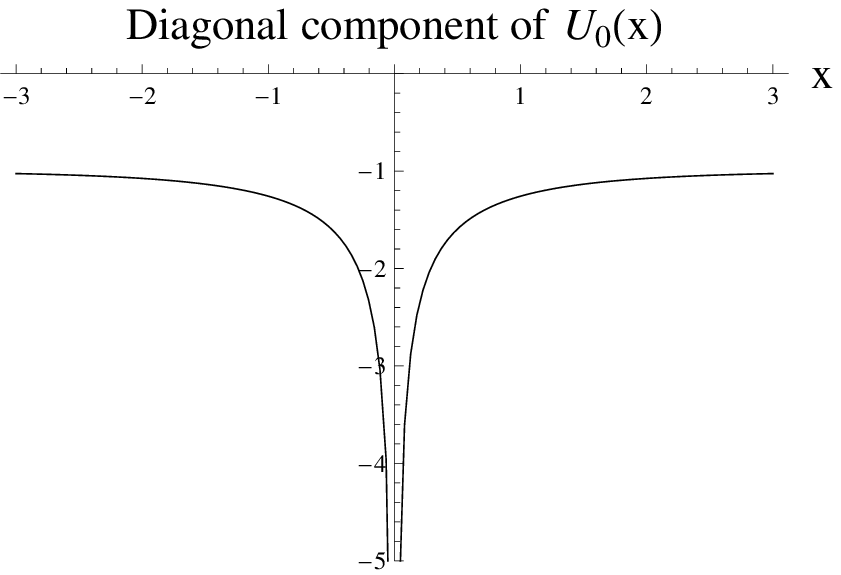,width=7.8cm}
\epsfig{file=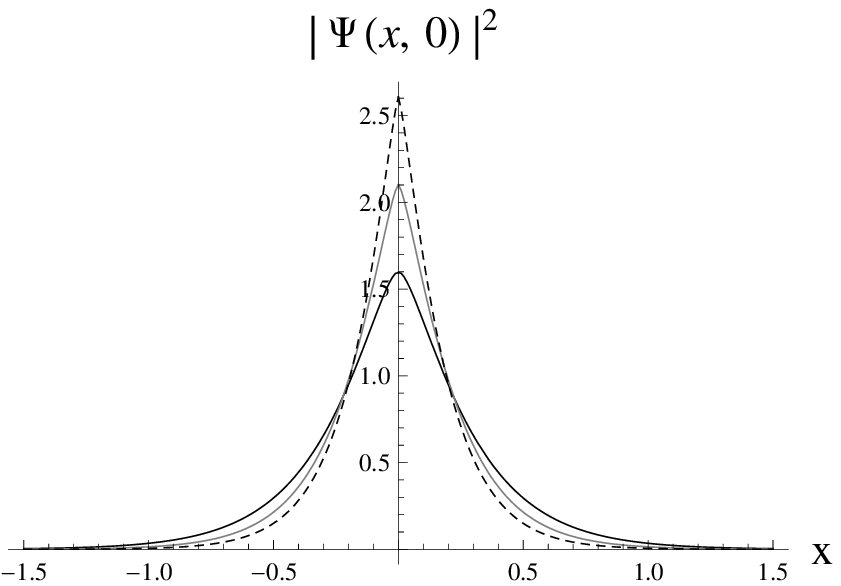,width=7.8cm}
\caption{Left plot: graph of the initial potential (\ref{u0boundexp}). Right plot: graphs of normalized 
probability densities associated with (\ref{psi1exp}), (\ref{psi2}) for the parameter settings (\ref{setx0exp}). We have 
$k_y=5/2$ (black solid curve), $k_y=3$ (gray curve), $k_y=7/2$ (dashed curve).}
\label{boundiniexp}
\end{center}
\end{figure}

\begin{figure}[h]
\begin{center}
\epsfig{file=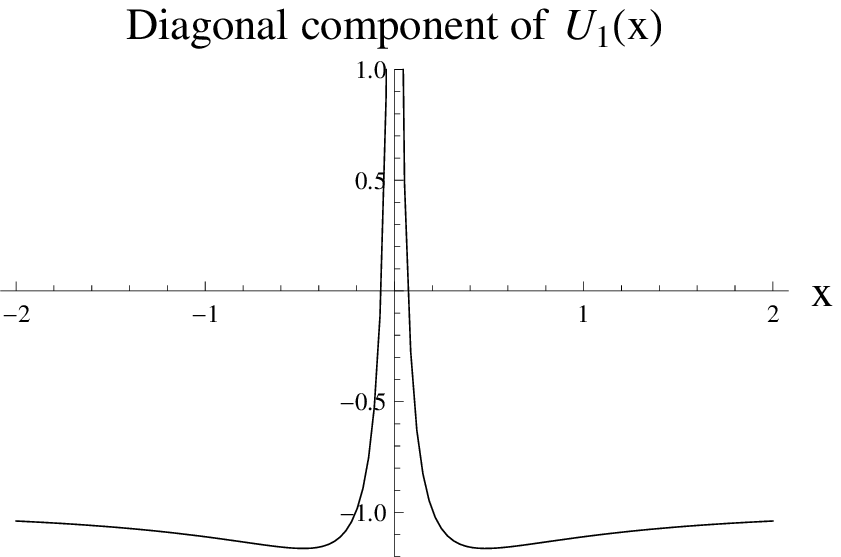,width=7.8cm}
\epsfig{file=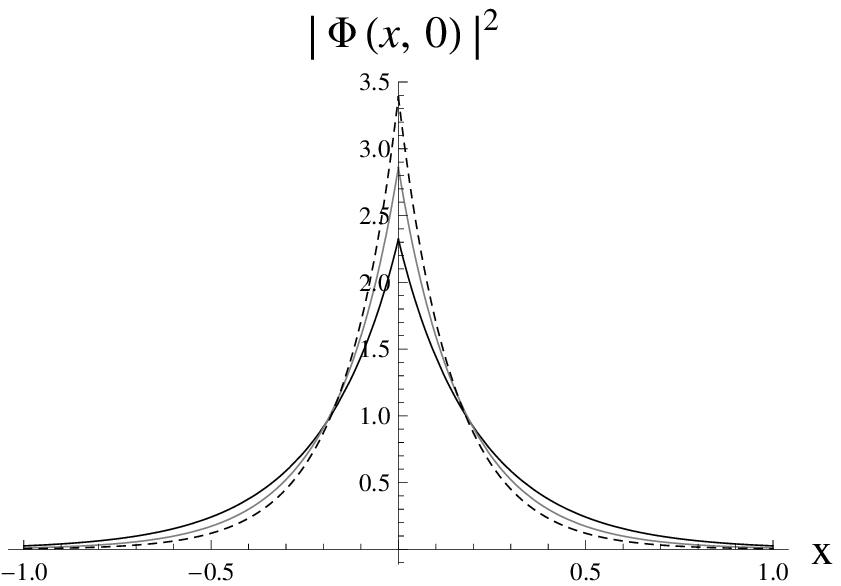,width=7.8cm}
\caption{Left plot: graphs of the transformed potential (\ref{v1}) for parameter settings (\ref{setx0exp}) and 
$\lambda_0=-\lambda_1=2$. Right plot: graphs of normalized 
probability densities associated with (\ref{psi1exp}), (\ref{psi2}) for the parameter settings (\ref{setx0exp}) 
and $\lambda_0=-\lambda_1=2$. We have 
$k_y=5/2$ (black solid curve), $k_y=3$ (gray curve), $k_y=7/2$ (dashed curve).}
\label{boundfinexp}
\end{center}
\end{figure}

\paragraph{Example: singular potential admitting bound states.} Similar to the system considered in section 3, we 
will now present an example that involves bound states. We will use the following parameter settings
\begin{eqnarray}
V_0~=~0 \qquad \qquad V_1=-1\qquad \qquad \sigma~=~-1 \qquad \qquad x_0 ~=~ -i~\pi. 
\label{setx0exp}
\end{eqnarray}
As outlined in the beginning of this section, the settings (\ref{setx0exp}) render the diagonal component of our 
potential (\ref{potexp}) as a function defined on the positive real axes with a singularity at the origin. We will now 
extend the domain of the present problem by replacing $x \rightarrow |x|$ in the potential. Upon taking into account 
our parameters (\ref{setx0exp}), we obtain
\begin{eqnarray}
U_0(x) &=&- \left[\frac{1}{\sqrt{1-\exp\left(-|x|\right)}} \right] I_2. \label{u0boundexp}
\end{eqnarray} 
A graph of the potential's diagonal component is visualized in the left plot of figure \ref{boundiniexp}.
 \noindent
The associated solutions (\ref{psi1exp}) and (\ref{psi2}) of our Dirac equation (\ref{dirac}) are constructed by substitution 
of (\ref{setx0exp}). It turns out that these solutions take elementary form if $k_y$ is an integer of absolute 
value greater than one: insertion of our parameter setting (\ref{setx0exp}) into the condition (\ref{gammacon1}) and 
solving for $k_y$ yields 
\begin{eqnarray}
|k_y| &=& \frac{n+1}{2}. \label{kybound}
\end{eqnarray}
The right plot in figure \ref{boundiniexp} shows graphs of the probability density $|\Psi|^2=|\Psi_1|^2+|\Psi_2|^2$ 
generated by our solutions. We observe that these densities are finite at zero and integrable at the infinities, such 
that they represent bound states of our system. As an example, let us state the explicit form of the solution 
(\ref{psi1exp}) for the value $k_y=5/2$:
\begin{eqnarray}
\Psi_1(x) &=& -\frac{\exp(-2 |x|)}{56~(-60+13~\sqrt{21})} 
\left[\frac{1}{2}-\frac{\sqrt{1-\exp(-|x|)}}{2}\right]^{\sqrt{21}}
\left[-1+\sqrt{1-\exp(-|x|)}\right]^{-\frac{\sqrt{21}}{2}} \times \nonumber \\[1ex] 
&\times& 
\left[1+\sqrt{1+\exp(-|x|)}\right]^{-\frac{\sqrt{21}}{2}} 
\Bigg\{1323-14 \sqrt{21}~i+63	\sqrt{21-21~\exp(-|x|)}+ \nonumber \\[1ex]
&+& \exp(|x|)  \Bigg[
-4704+142 \sqrt{21}~i- 
644~\sqrt{21-21~\exp(-|x|)}\Bigg]+ \nonumber \\[1ex]
&+&\exp(2~|x|)~
\Bigg[3500-160 \sqrt{21}~i+764~\sqrt{21-21~\exp(-|x|)}-728~i~\sqrt{1-\exp(-|x|)}\Bigg]
\Bigg\}. \nonumber
\end{eqnarray}
In the next step we perform our Darboux transformation, choosing the transformation parameters as 
$\lambda_0=2=-\lambda_1$. As we are looking to generate bound states, we can use the argumentation 
presented in the case originating from the potential (\ref{u0bound}): if the initial solution components and the entries of 
our transformation matrix vanish at the infinities, so will the transformed solution components. Now, 
since this value solves our condition (\ref{kybound}) for $n=1$, our transformation 
matrix (\ref{u}) is elementary, and so is the transformed potential (\ref{v1}). More precisely, the latter potential 
is given in explicit form by
\begin{eqnarray}
U_1(x) &=& \Bigg\{
-4725+420~\exp(|x|)~\Bigg[274+45~\sqrt{3-3\exp(-|x|)}\Bigg]- \nonumber \\[1ex]
&-& 128~\exp(4|x|)~
\Bigg[1933+1116~\sqrt{3-3~\exp(-|x|)}\Bigg] \nonumber \\[1ex]
&-& 120~\exp(2|x|)~
\Bigg[3803+1211~\sqrt{3-3~\exp(-|x|)}\Bigg]+ \nonumber \\[1ex]
&+& \exp(3|x|)~
\Bigg[594816+272000~\sqrt{3-3~\exp(-|x|)}\Bigg]\Bigg\}  \times \nonumber \\[1ex]
&\times&
\Bigg\{
3675~\Bigg[4~\sqrt{3}+\sqrt{1-\exp(-|x|)}\Bigg]- \nonumber \\[1ex]
&-&100~\exp(|x|)~
\Bigg[1321~\sqrt{3}+906~\sqrt{1-\exp(-|x|)}\Bigg]- \nonumber \\[1ex]
&-&128~\exp(3|x|)~
\Bigg[2991~\sqrt{3}+4214~\sqrt{1-\exp(-|x|)}\Bigg]+ \nonumber \\[1ex]
&+&40~\exp(2|x|)~
\Bigg[8935~\sqrt{3}+9507~\sqrt{1-\exp(-|x|)}\Bigg]+ \nonumber \\[1ex]
&+&\exp(4|x|)~
\Bigg[142848~\sqrt{3}+247424~\sqrt{1-\exp(-|x|)}\Bigg]
\Bigg\}^{-1} I_2. \nonumber
\end{eqnarray}
As can be seen from the latter expression, the transformed 
potential matrix is real-valued and diagonal, since our parameter settings satisfy conditions (\ref{real5}) 
and (\ref{real6}). In the last step we substitute the parameter settings into (\ref{darboux}), 
obtaining transformed Dirac solutions, the explicit form 
of which we do not show here due to their length. Instead, we visualize these quantities in figure \ref{boundfinexp}.
 \noindent
Note that the probability densities shown in the figure are smooth at the origin.

\section{Concluding remarks} 
In this work we have applied our first-order Darboux transformation to the massless Dirac equation for two 
classes of potentials admitting hypergeometric solutions. Since our Darboux transformation preserves 
boundary conditions, provided the transformation parameters are chosen appropriately, we were able to 
generate a variety of Dirac solutions with associated potentials, including the case of bound states. A 
particular focus was put on the construction of elementary solutions through the Darboux transformation. 
Note that the property of a solution to be elementary does not imply a specific behavior. For example, 
the bound states found from potential (\ref{u0bound}) are not elementary functions, while their counterparts 
generated from (\ref{u0boundexp}) are elementary. Our results can be 
extended in several ways, one of which is application of our methods to different Dirac systems. A further way 
of generalization is to consider higher-order Darboux transformations. Unfortunately, to the best of our knowledge there 
is no concise form of such transformations, as there is in the Schr\"odinger case through Wronskian 
representations. It will be subject to future research, as well as the search for elementary cases and 
Darboux partners of massless Dirac equations that admit closed-form solutions.

\end{sloppypar}


\begin{thebibliography}{99}

\bibitem{abram} M. Abramowitz and I.A. Stegun,
$"$Handbook of Mathematical Functions with Formulas, Graphs, and Mathematical Tables$"$,
Dover, 1964

\bibitem{allain} P.E. Allain and J.N. Fuchs, $"$Klein tunneling in graphene: optics with massless electrons$"$, 
Eur. Phys. J. B 83 (2011), 301 

\bibitem{bagrov} V.G. Bagrov and B.F. Samsonov,  
$"$Supersymmetry of a nonstationary Schr\"odinger equation$"$, 
Phys. Lett. A 210 (1996), 60-64.

\bibitem{bala} A.V. Balatsky, I. Vekhter, and J.-X. Zhu, 
$"$Impurity-induced states in conventional and unconventional superconductors$"$, 
Rev. Mod. Phys. 78 (2006), 373


\bibitem{cayssol} J. Cayssol, $"$Introduction to Dirac materials and topological insulators$"$, Comptes Rendus
Physique 14 (2013), 760


\bibitem{darboux} G. Darboux, $"$Sur une proposition relative aux \'equations lin\'eaires$"$, C. R. Acad. Sci. 94 (1882), 
1456

\bibitem{downingzero} C. A. Downing and M. E. Portnoi, $"$Zero-energy vortices in Dirac materials$"$, Phys. Status
Solidi B 256 (2019), 1800584

\bibitem{radial} C.A. Downing, A.R. Pearce, R.J. Churchill, and M.E. Portnoi, $"$Optimal traps in graphene$"$, 
Phys. Rev. B 92 (2015), 165401

\bibitem{down0} C. A. Downing, D.A. Stone, and M.E. Portnoi, $"$Zero-energy states in graphene quantum dots and rings$"$,
Phys. Rev. B 84 (2011), 155437


\bibitem{erementchouk} M. Erementchouk, P. Mazumder, M.A. Khan, and M.N. Leuenberger, $"$Dirac electrons
in the presence of matrix potential barrier: application to graphene and topological
insulators$"$, J. Phys. Condens. Matter 28 (2016), 115501


\bibitem{djtrend} D.J. Fernandez, $"$Trends in Supersymmetric Quantum Mechanics$"$. 
In: Kuru S., Negro J., Nieto L. (eds) Integrability, Supersymmetry and Coherent States. 
CRM Series in Mathematical Physics. Springer, Cham, 2019

\bibitem{djsusy} D.J. Fernandez, $"$Supersymmetric quantum mechanics$"$,
AIP Conf. Proc. 1287 (2010), 3-36

\bibitem{geim} A.K. Geim and K.S. Novoselov, $"$The rise of graphene$"$, Nature Materials 6 (2007), 183

\bibitem{gu} C. Gu, A. Hu, and Z. Zhou, $"$Darboux Transformations in Integrable Systems$"$, (Springer 
Science and Business Media, Dordrecht, 2005)

\bibitem{hartmann3} R.R. Hartmann, N.J. Robinson, and M.E. Portnoi, 
$"$Smooth electron waveguides in graphene$"$, Phys. Rev. B 81 (2010), 245431

\bibitem{arturexp} A. Ishkhanyan, $"$Exact solution of the Schr\"odinger equation for a short-range 
exponential potential with inverse square root singularity$"$, Eur. Phys. J. Plus 133 (2018), article number 83

\bibitem{artur} A. Ishkhanyan, $"$A singular Lambert-W Schr\"odinger potential exactly solvable in terms of the confluent
hypergeometric functions$"$, Mod. Phys. Lett. A 31 (2016), 165

\bibitem{artur2} A. Ishkhanyan, $"$The Lambert-W step-potential - an exactly solvable confluent
hypergeometric potential$"$, Phys. Lett. A 380 (2016), 640

\bibitem{arturepl} A.M. Ishkhanyan, $"$The third exactly solvable hypergeometric quantum-mechanical 
potential$"$, EPL 115 (2016), 20002 

\bibitem{jakubsky} V. Jakubsky, $"$Spectrally isomorphic Dirac systems: graphene in electromagnetic field$"$,
Phys. Rev. D 91 (2015), 045039

\bibitem{kats} M.I. Katsnelson, K.S. Novoselov, and A.K. Geim, 
$"$Chiral tunnelling and the Klein paradox in graphene$"$, Nat. Phys.2 (2006), 620

\bibitem{matveev} V.B. Matveev and M.A. Salle,  
$"$Darboux transformations and solitons$"$, (Springer Science and Business Media, Berlin, 1991)

\bibitem{nieto} L.M. Nieto, A.A. Pecheritsin, and B.F. Samsonov, $"$Intertwining technique for the one-dimensional stationary
Dirac equation$"$, Ann. Phys. 305 (2003), 151

\bibitem{ekat} E. Pozdeeva and A. Schulze-Halberg, $"$Darboux transformations for a generalized 
Dirac equation in two dimensions$"$, J. Math. Phys. 51 (2010), 113501

\bibitem{wehling} T.O. Wehling, A.M. Black-Schaffer, and A.V. Balatsky, 
$"$Dirac materials$"$,  Advances in Physics 63 (2014), 1


\end{thebibliography}
\end{document}